\documentclass[journal]{IEEEtran}
\ifCLASSINFOpdf
\else
\fi

\IEEEoverridecommandlockouts
\usepackage{epsfig}
\usepackage{color}
\usepackage{amssymb,latexsym,amsfonts,amsmath,mathrsfs}
\usepackage{graphicx}
\usepackage{algorithm} 
\usepackage{algorithmic}
\usepackage{cite}
\usepackage{bbm}
\usepackage{booktabs}
\usepackage{tabularx}
\allowdisplaybreaks[4]

\newtheorem{thm}{Theorem}
\newtheorem{lem}{Lemma}
\newtheorem{prop}{Propsition}

\newtheorem{assum}{Assumption}
\newtheorem{rem}{Remark}

\begin{document}
%
\title{High Performance Distributed Control for Large-Scale Linear Systems: A Cover-Based Distributed Observer Approach}
%
%
%

\author{Haotian~Xu,
        Shuai~Liu,~\IEEEmembership{Member,~IEEE},
        Ling~Shi,~\IEEEmembership{Fellow,~IEEE},
\thanks{This work is supported by National Natural Science Foundation of China (No.61533013, 61633019, 61903290), Funded by China Postdoctoral Science Foundation (No.2022M721932). The corresponding author: Shuai~Liu.}
\thanks{H. Xu is with Department of Automation, Shanghai Jiao Tong University, Key Laboratory of System Control and Information Processing, Ministry of Education of China, Shanghai, 200240~(e-mail: xuhaotian\_1993@126.com; xuhaotian@sjtu.edu.cn)}
\thanks{L. Shi is with Department of Electronic and Computer Engineering, The Hong Kong University of Science and Technology, Hong Kong, China (e-mail: eesling@ust.hk; wshchen@126.com).}}

%
%

\markboth{Submitted to IEEE Transactions on Automatic Control}%
{Haotian Xu \MakeLowercase{\textit{et al.}}: Distributed Observer Design}
%



\maketitle

\begin{abstract}
In recent years, the distributed-observer-based distributed control law has shown powerful ability to arbitrarily approximate the centralized control performance. However, the traditional distributed observer requires each local observer to reconstruct the state information of the whole system, which is unrealistic for large-scale scenarios. To fill this gap, This paper presents a coverage solution algorithm for large-scale systems that accounts for both physical and communication network characteristics, which can significantly reduce the dimension of local observers. Then, the cover-based distributed observer for large-scale systems is proposed to overcome the problem that the system dynamics are difficult to estimate due to the coupling between cover sets. Furthermore, the two-layer Lyapunov analysis method is adopted and the dynamic transformation lemma of compact errors is proved, which solves the problem of analyzing stability of the error dynamic of the cover-based distributed observer. Finally, it is proved that the distributed control law based on the cover-based distributed observer can also arbitrarily approximate the control performance of the centralized control law, and the dimension of the local observer is greatly reduced compared with the traditional method. The simulation results show the validity of the developed theories.
\end{abstract}


%
\IEEEpeerreviewmaketitle

\section{Introduction}\label{sec1}
A distributed observer comprises a network of 
$N$ cooperative agents connected via a communication graph. Each agent acts as a node, equipped with a local observer and a local controller. While each local observer only accesses partial system outputs, they collaborate through the communication network to collectively estimate the entire system's state. Due to their broad applicability—demonstrated in fields such as flexible structures \cite{Zhang2018Distributed}, smart vehicles \cite{Xu2022TIV}, microgrids \cite{6727407,9305705}, and spacecraft \cite{8118289}—and significant theoretical value, distributed observers have attracted considerable research interest \cite{battilotti2019distributed,2019Completely,deghat2019detection,Han2017A,Xu2021IJRNC,9461598,8093658,Liu2018Cooperative,YANG2023110690,9829291,Xu2022Cybernetics,wang2024split,baum2024distributed,duan2024framework}. Theoretically, they enable distributed control laws to achieve performance levels that can arbitrarily approximate centralized control within a sufficiently small error \cite{Xu2022TIV,XU2023ISA,Xu2020Distributed,8985536,2021DistributedMeng}, a key advancement in the field.

However, current distributed observer designs suffer from a critical limitation: they require each agent to reconstruct the full state vector of the entire system. This requirement, known as state omniscience, implies that if the global system dimension is 
$n$, every local observer must also have dimension $n$. For large-scale systems, this becomes computationally prohibitive, contradicting the fundamental goal of distributed design. Although some studies have explored minimum-order \cite{Han2018Towards} or reduced-order observers \cite{wang2022Dis}, they typically replace $N$ full-order observers with $N$ lower-order observers. {\color{blue}Since the dimension $p_i$ of the measurement output of each local observer is much smaller than that of the large-scale system state, the distributed reduced-order observers does not fundamentally solve the scalability issue caused by state omniscience (the dimension of each local observer is $n-p_i$).}

This paper investigates whether the dimension of local observers can be significantly reduced. The central challenge is to achieve this reduction without compromising the ability of the resulting distributed control law to approach centralized performance. Existing work \cite{ZECEVIC2005265} suggests that the performance gap between distributed and centralized control stems from the former's limited access to state information. Conversely, distributed observers bridge this gap precisely through state omniscience \cite{Xu2022TIV}, indicating that state omniscience is a key factor in enabling distributed-observer-based distributed control laws to approximate centralized performance. Therefore, replicating this performance without state omniscience presents a major challenge.

Our approach is inspired by insights from the structure of large-scale interconnected systems. We observe that for such systems—often represented by a physical network graph—the local control law on an agent typically does not require the full states of the entire system. {\color{cyan}Theoretical analyses demonstrate that by truncating feedback connections beyond a certain topological distance in the physical network, the performance degradation relative to the centralized optimal control is exponentially small. Therefore, the states of its physical network neighbors are sufficient to achieve performance that closely approximates the centralized benchmark.} This insight leads to our core question: Can we design a novel distributed observer where each local observer only estimates the states of its physical neighbors? If feasible, the local observer dimension would be drastically reduced, especially since large-scale systems are often sparsely connected. Moreover, if an agent can estimate all its physical neighbors' states, its local controller can utilize similar information as a centralized one, preserving performance while reducing observer dimension.

Realizing this novel observer is highly non-trivial. A local observer can estimate its physical neighbors' states directly only if the physical and communication networks are identical---{\color{blue}a condition rarely met in practice (For example, smart grid systems, urban water supply systems, etc., their physical networks need to reach thousands of households, but the communication networks responsible for data transmission and control signal issuance can be more flexible and sparse)}. This mismatch means that a single agent's communication neighbors may not cover its physical neighbors. Consequently, this paper develops a node covering method where each cover set contains multiple nodes, and nodes can belong to multiple cover sets. A distributed observer is then designed within each cover set. The key is to ensure that for each agent, the union of the cover sets it belongs to covers all its physical neighbors (a necessary condition for approximating centralized performance), while minimizing the total number of involved nodes to reduce local observer dimension.

This leads to the first scientific challenge: designing an intelligent cover solving strategy that meets the above conditions. {\color{blue}The literature on coverage or similar partitioning primarily addresses schemes that either divide networks into non-overlapping sections based on specific tasks or requirements \cite{4653485,7859409}, optimize partitions against defined objectives \cite{9524445,2001Algorithms,7502076,9112722}, or coordinate agents for complete areal coverage \cite{7934022,9458547}.} {\color{red}While the third category aligns with our goal, existing methods presuppose known agent parameters and areal coverage, unlike our problem of covering network nodes. Furthermore, in our scenario, both the regions to be covered and the number of covering sets are unknown and must be co-determined through an analysis of the physical and communication networks. Consequently, this divergence renders the literature inadequate for our needs, thus requiring a new algorithmic approach.}



The second challenge involves managing the coupled error dynamics across cover sets. Although observers are designed within cover set, their dynamics are coupled with observers in other cover sets. This coupling must be carefully handled in both the observer design and stability analysis. If this coupling relationship is not eliminated, then a distributed observer within the cover sets cannot be designed. Whereas, if the coupling is directly eliminated, then the adverse impacts caused by issues such as model mismatch {\color{blue}(there are differences between the model used by the observer and the actual model of the system)} must be considered. rendering classical distributed observer theories inadequate for this new architecture.

To address these challenges, this paper makes three key contributions: First, we propose a novel and feasible coverage solving method for large-scale systems, which yields a Pareto-optimal solution for the multi-objective optimization problem inherent in each agent's conflicting preferences for cover set selection. Second, we present a design method for the cover-based distributed observer and develop a two-layer Lyapunov approach to analyze its error dynamics' stability. Finally, we demonstrate that the distributed control law based on our cover-based distributed observer can arbitrarily approximate centralized control performance within a sufficiently small error bound, {\color{cyan}with the performance gap proven to be exponentially small relative to the truncation distance in the physical network topology}, providing a theoretical foundation for applying distributed observers to large-scale systems.

The rest of this paper is organized as follows. Section \ref{sec2} formulates the problem. Section \ref{sec3} details the coverage solving method and observer design. Section \ref{sec4} analyzes the error dynamics and closed-loop performance. Simulation results are presented in Section \ref{sec5}, and Section \ref{sec6} concludes the paper.

\textit{Notations:~}Let $\mathbb{R}^n$ be the Euclidean space of $n$-dimensional column vectors and $\mathbb{R}^{m\times n}$ be the Euclidean space of $m\times n$-dimensional matrices. Denote $A^T$ the transpose of matrix $A$. $\bar{\sigma}(A)$ and $\underline{\sigma}(A)$ stand for the maximum and minimum eigenvalues of $A$, respectively, if $A=A^T$. We cast $col\{A_1,\ldots,A_n\}$ as $[A_1^T,\ldots,A_n^T]^T$, and $diag\{A_1,\ldots,A_n\}$ as a block diagonal matrix with $A_i$ on its diagonal, where $A_1,\ldots,A_n$ are matrices with arbitrary dimensions. $dim\{\cdot\}$ represents the dimension of a vector. $I_n\in\mathbb{R}^{n\times n}$ is an identity matrix and $sym\{A\}=A+A^T$ if $A$ is a square matrix. We denote $|\cdot|$ the cardinality of a set. $\|\cdot\|$ represents the $2$-norm of vectors or matrices. Furthermore, there are many symbols in this paper. To avoid confusion for readers, we have listed a comparison table of easily confused symbols in Table 1, and corresponding symbols will also be explained when they first appear.
\begin{table}[!h]
	\label{table}
	\caption{Symbol description}
	\centering
	\renewcommand\arraystretch{1.3}
	\begin{tabular}{p{1cm}<{\centering}p{7cm}<{\raggedright}}
		\hline
		Symbol & Definition\\
		\hline
		$\mathcal{O}_p$ & The $p$th cover set \\
		$\mathscr{P}_i$ & Set of cover sets that containing node $i$ \\
		$\mathcal{O}(\mathscr{P}_i)$ & $\bigcup_{\mathcal{O}_j\in\mathscr{P}_i}\mathcal{O}_j$\\
		$D(\mathscr{P}_i)$ & ${\color{blue}\sum_{\mathcal{O}_j\in\mathscr{P}_i}|\mathcal{O}_j|}$ \\
		$\hat{x}_{li}^{(p)}$    &  Estimation of $x_i$ generated by observers on agent $l$ in cover set $\mathcal{O}_p$\\
		$\bar{x}_{li}$    & $\bar{x}_{li}=1/N_{i,\mathscr{P}_l}\cdot\sum_{\mathcal{O}_p\in\mathscr{P}_l}\hat{x}_{li}^{(p)}$ \\
		$\mathbbm{x}_{li}$ & Saturate value of $\bar{x}_{li}$\\
		$e_{\star i}^{(p)}$ & $col\{e_{ji}^{(p)}\},~j\in\mathcal{O}_p$\\
		$e_{i \star}$ & $col\{e_{ij}^{(p)},~j\in\mathcal{O}_p,~\mathcal{O}_p\in\mathscr{P}_i\}$\\
		\hline    
	\end{tabular}
\end{table}

\section{Problem formulation}\label{sec2}

\subsection{Overall objectives}\label{sec2.1}

Consider a large-scale interconnected system composed by $N$ subsystems (corresponding to $N$ agents) and the $i$th subsystem takes the form of:
\begin{align}
	&\dot{x}_i=\sum_{j=1}^NA_{ij}x_j+B_iu_i,\label{sys11}\\
	&y_i=C_ix_i,\label{sys12}
\end{align} 
where $x_i\in\mathbb{R}^n$, $y_i\in\mathbb{R}^p$, and $u_i\in\mathbb{R}^m$ are the system states, output measurements, and control inputs of the $i$th subsystem, respectively; $x_j\in\mathbb{R}^n$ is the state of the $j$th subsystem; The matrices $A_{ii}$, $B_i$, and $C_i$ represent the system matrix, control matrix, and output matrix of the $i$th subsystem with compatible dimensions, respectively; and $A_{ij}\in\mathbb{R}^{n\times n}$ stands for the coupling matrix between the $i$th and the $j$th subsystems. Let $A=[A_{ij}]_{i,j=1}^N$, $B=diag\{B_1,\ldots,B_N\}$, and $C=col\{C_1,\ldots,C_N\}$, then the compact form of the large-scale system is given by
\begin{align}
	&\dot{x}=Ax+Bu,\label{sys21}\\
	&y=Cx,\label{sys22}
\end{align}
where $x=col\{x_1,\ldots,x_N\}\in\mathbb{R}^{nN}$, $y=col\{y_1,\ldots,y_N\}\in\mathbb{R}^{pN}$, and $u=col\{u_1,\ldots,u_N\}\in\mathbb{R}^{mN}$. 

We construct a physical network for this large-scale system by its interconnection. Let $\mathcal{V}$ be the nodes (agents) set and we say $(j,i)$ is an edge connecting $j$ and $i$ if $\|A_{ij}\|\neq 0$. Then, the set $\mathcal{E}_p$ is defined by $\mathcal{E}_p=\{(i,j):~\|A_{ji}\|\neq 0\}$. Accordingly, the adjacency matrix is defined as $\mathcal{A}_p=[\beta_{ij}]_{i,j=1}^N$ where $\beta_{ij}=1$ if $(j,i)\in\mathcal{E}_p$. Let $\mathcal{N}_i=\{j:~\beta_{ij}=1\}$ be the set of physical network neighbors of agent $i$.
The physical network can be directed or undirected and the schematic diagrams in this paper all use undirected graphs, but it should be noted that when the physical network is an undirected graph, the system matrix is assumed to have constraints with $\|A_{ij}\|=\|A_{ji}\|=0$; When the physical network is a directed graph, the system matrix has no constraints. Moreover, an undirected graph $\mathcal{G}_c=\{\mathcal{V},\mathcal{E}_c,\mathcal{A}_c\}$ represents the communication network among all nodes, where $\mathcal{E}_c$ and $\mathcal{A}_c=[\alpha_{ij}]_{i,j=1}^N$ are the associated edge set and adjacency matrix, respectively. Then, the set of communication network neighbors of agent $i$ is given by $\mathcal{C}_i=\{j:~\alpha_{ij}=1\}$. The Laplacian matrix associated to $\mathcal{G}_c$ is denoted by $\mathcal{L}_c=\mathcal{D}_c-\mathcal{A}_c$, where $\mathcal{D}_c=diag\{d_i,~i=1,\ldots,N\}$ with $d_i=-\alpha_{ii}+\sum_{j=1}^N\alpha_{ij}$.

\begin{figure}[!t]
	\centering
	\includegraphics[width=8cm]{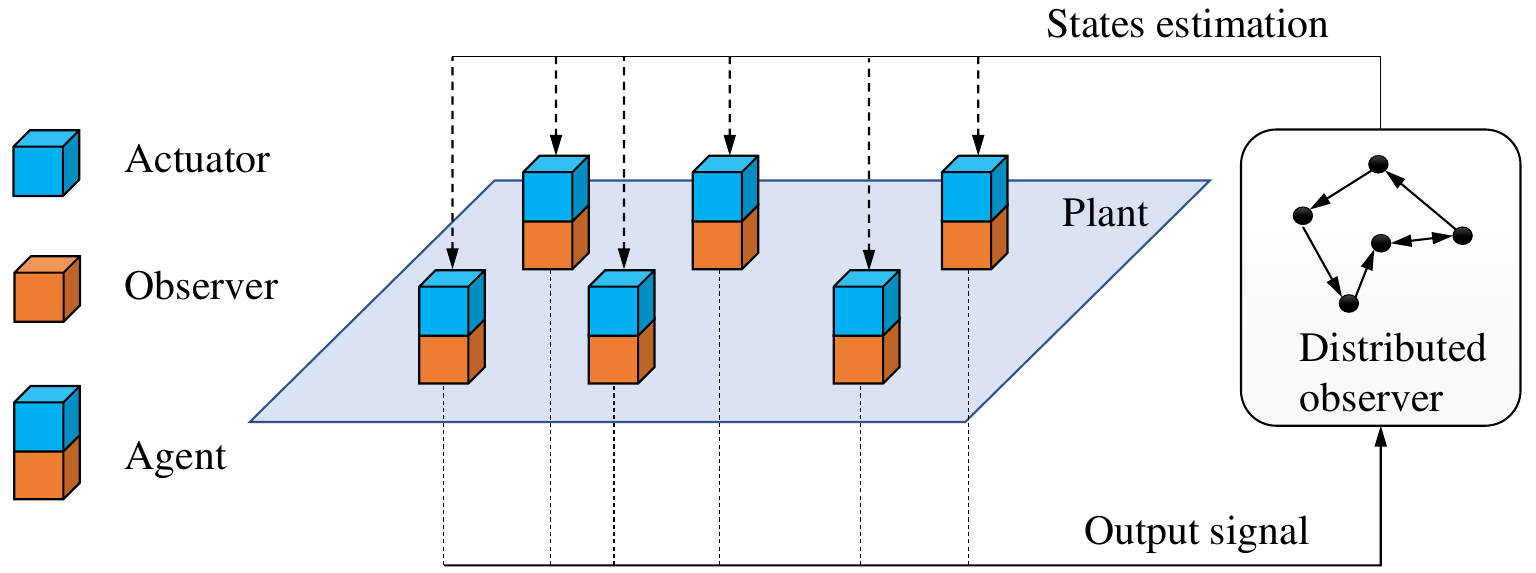}\\
	\caption{Architecture diagram of distributed observer.}\label{observer-tu}
\end{figure}

In traditional distributed observer, each local observer---the orange block in Figure \ref{observer-tu}---is required to reconstruct all the states of the entire system based on $y_i$ and the information exchanged via communication network. Then, each local controller---blue block in Figure \ref{observer-tu}---can use all the information of the entire system. Therefore, all local observers can jointly establish a distributed control law with centralized performance \cite{Xu2022TIV,Xu2020Distributed}. This paper will investigate whether it is possible to achieve distributed control laws with arbitrary approximation of centralized control performance without requiring each local observer to estimate global information.


To formulate the problem, we assume that there is a centralized control law $u(x)=col\{u_1(x),\ldots,u_N(x)\}$ such that system (\ref{sys21}) is stabilized, and denote $x_c(t)$ the solution of (\ref{sys21}) with centralized control law. We further assume a distributed control law $\bar{u}_i(\hat{x}_{i\star})$ such that $x_r(t)$ solved by (\ref{sys21}) with $u=\bar{u}(\hat{x}_{i\star})$ is stabilized, where $\hat{x}_{i\star}$ is the states estimation generated by the $i$th local observer located at the $i$th agent, and $\bar{u}(\hat{x}_{i\star})=col\{\bar{u}_i(\hat{x}_{i\star}),~i=1,\ldots,N\}$. Subsequently, the overall goal of this paper consists of the following two parts. 

1)~Design the cover-based distributed observer with $\hat{x}_{i\star}$ being the states of the $i$th local observer, and $\hat{x}_{i\star}$ satisfies $dim\{\hat{x}_{i\star}\}\ll dim\{x\}$;

2)~Design a distributed-observer-based distributed control law $\bar{u}(\hat{x}_{i\star})$ such that $x_r(t)$ can approach $x_c(t)$ arbitrarily, i.e., $\|x_c(t)-x_r(t)\|<\varepsilon$, $\forall\varepsilon>0$ and $\forall t\geq 0$.

\subsection{An illustrative example for the target problem}\label{sec2.2}

\begin{figure*}[!t]
	\centering
	\includegraphics[width=18cm]{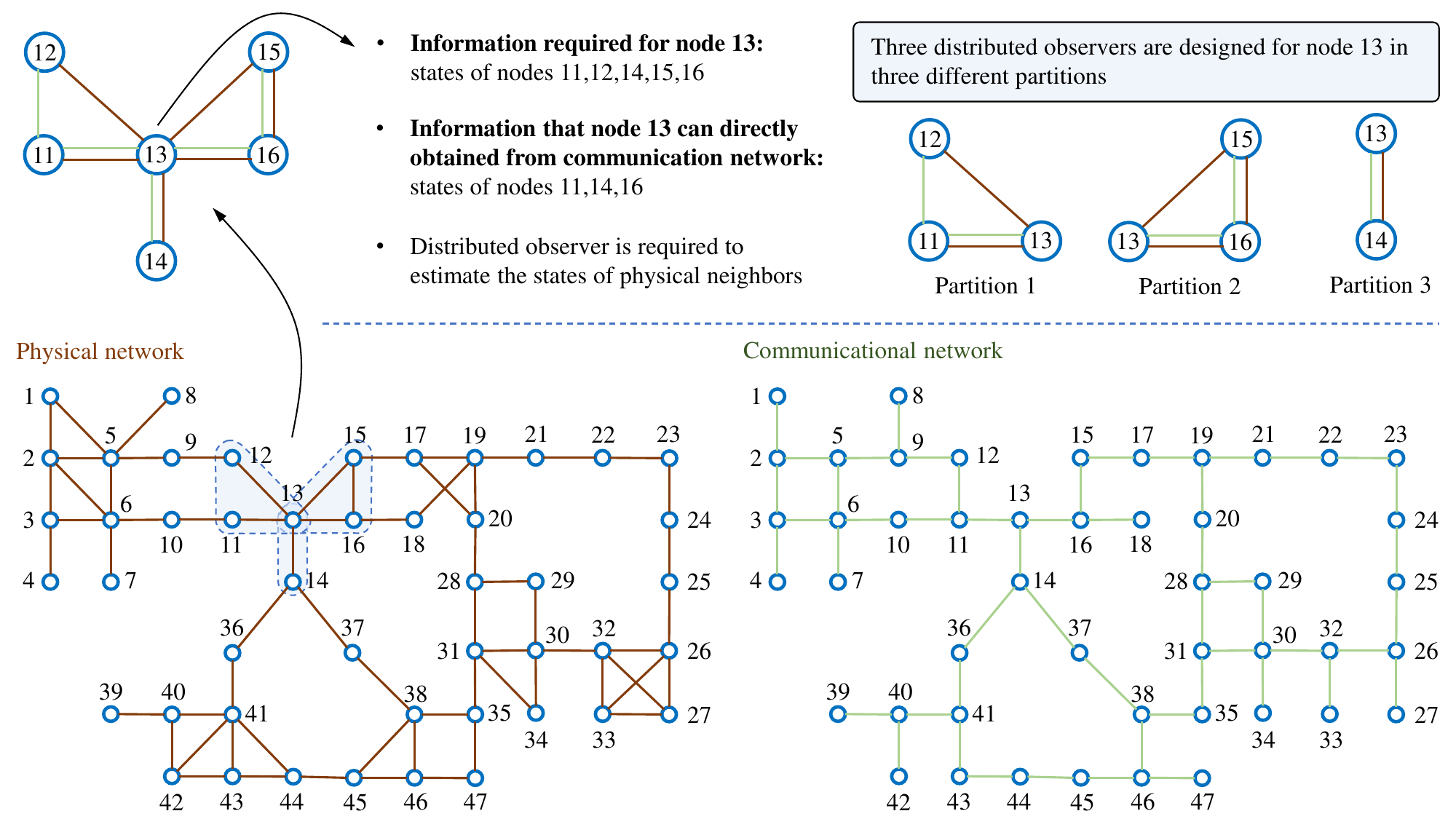}\\
	\caption{Illustrations of coverage issue in simple network systems.}\label{target}
\end{figure*}

In this subsection, we will illustrate an example for helping readers understand the problems proposed in the above subsection. See in Figure \ref{target}, the coupling relationship of a large-scale system is shown in the physical network, in which each node represents an subsystem. Assume that each subsystem has a corresponding agent, and the communication relationship among all agents is shown as communication network. The $i$th agent has access to the information of $y_i$, and is supposed to establish several local observers and a local control law.

We take node $13$ as an example to demonstrate the specific issues to be studied in this paper. As seen in Figure \ref{target}, node $13$ has $5$ physical neighbors ($11,12,14,15,16$) and $3$ communication neighbors ($11,14,16$). To design a local control law to make the control performance as close as possible to the centralized control law, agent $13$ needs the states of all its physical neighbors. In traditional method, a distributed observer will be designed such that agent $13$ can reconstruct all states of the entire system. However, only the states of $11,12,14,15,16$ are required. Therefore, this paper intends to design a group of cover sets and design distributed observer in each cover set. In this way, the useless information estimated by each agent can be greatly reduced. For example, in Figure \ref{target}, node $13$ belongs to $3$ cover sets---$\{11,12,13\}$, $\{13,15,16\}$, and $\{13,14\}$. By designing distributed observers separately in three cover sets, agent $13$ only needs to estimate the states of $5$ subsystems (in traditional methods, agent $13$ needs to estimate the states of all $47$ subsystems). With cover-based distributed observers, local control laws can be established based on the results of state estimation.

Up to now, we have briefly described the main idea of this paper through an example. To achieve this idea, there are mainly three steps involved: 1) Design a reasonable and effective coverage solving algorithm; 2) Design cover-based distributed observer in each cover set; 3) Design distributed control law.


Before achieving these steps, we first specify two basic assumptions and an important lemma. 

\begin{assum}\label{assume1}
	Pair $(A,B)$ is assumed to be controllable and $(C_{i},A_{ii})$ for $i=1,\ldots,N$ are  assumed to be observable.
\end{assum}
\begin{assum}\label{assume2}
	Graphs corresponding to communication network is assumed to be undirected, connected and simple (A graph or network is simple means there is no self-loop and multiple edges). 
\end{assum}
\begin{lem}[\cite{Hong2008Distributed}]\label{spanning}
	Consider an undirected connected graph $\mathcal{G}=\{\mathcal{V},\mathcal{E},\mathcal{A}\}$. Let $\mathcal{S}=diag\{1,0,\ldots,0\}\in\mathbb{R}^{N\times N}$ and then all eigenvalues of matrix $\mathcal{H}=\mathcal{L}+\mathcal{S}$ locate at the open right half of plane, where $\mathcal{L}$ is the Laplacian matrix of $\mathcal{G}$.   
\end{lem} 
\begin{rem}
	The reader should note that Assumption \ref{assume1} is not completely the same as the traditional assumption in distributed observers. In existing literature \cite{Han2017A,Xu2020Distributed}, it is generally assumed that $(C,A)$ is observable, but $(C_i,A)$ may not be observable. Assumption \ref{assume1} used in this paper is given owing to the form of system (\ref{sys11})--(\ref{sys12}). This system format is easier to express our core views on covering and status information retrieval. Actually, the cover-based distributed observer can also be achieved based on the General system $\dot{x}=Ax,~y=Cx$ and the traditional assumption as in \cite{Han2017A,Xu2020Distributed}. However, due to limitations in space and the need to express the main idea, the theory of cover-based distributed observers for general systems will be included in our future research.
\end{rem}

\section{Design of cover-based distributed observer}\label{sec3}

A large-scale network coverage solving algorithm will be given in this section. Then, based on the coverage solving results, we propose the design method and observer structure of the cover-based distributed observer.

\subsection{Cover sets selecting}\label{sec3.1}
As mentioned in Section \ref{sec1} and Subsection \ref{sec2.2}, it is unnecessary to require each local observer to reconstruct the states of the whole large-scale system. Theoretically, each agent only needs to estimate part of states required by its local control law. Therefore, in order to establish a distributed observer that can only estimate the state information required by the local control law, we need to find a cover of all agents and ensure that all cover sets of each agent contain all its physical network neighbors {\color{blue}(note that in many large-scale systems such as power grids and water networks, communication networks are often distinct from physical networks. This is because communication infrastructure can be flexibly arranged to reduce costs, rather than being constrained by the complex pipeline layouts of physical systems)}. To illustrate it in more detail, we introduce several simple examples.


\begin{figure}[!t]
	\centering
	\includegraphics[width=8cm]{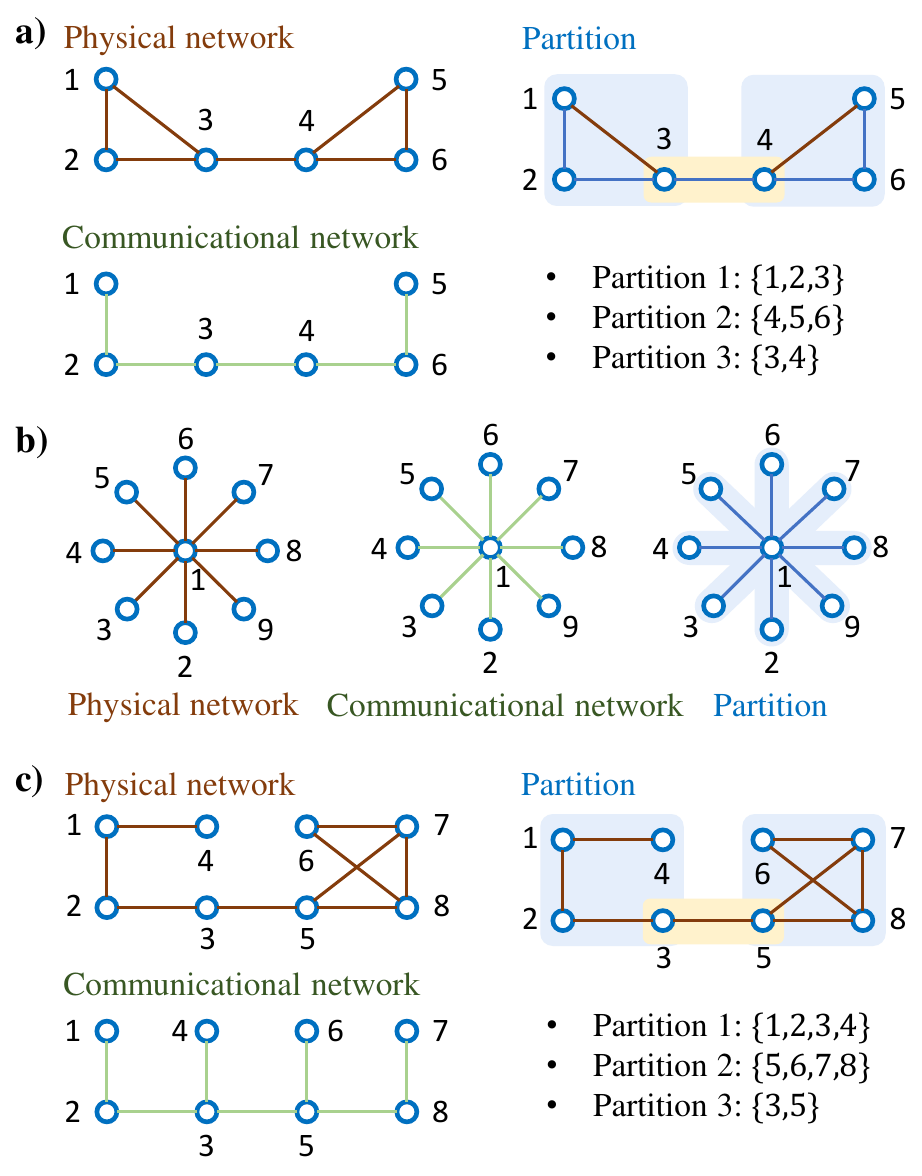}\\
	\caption{Illustrations of covering in simple network systems.}\label{partition}
\end{figure}

In Figure \ref{partition}a), there is a simple network system with $6$ nodes. One may recognize its cover by observation easily: all nodes could be covered by three sets including $\{1,2,3\}, \{4,5,6\}$, and $\{3,4\}$. With this cover sets, the physical network neighbors of all nodes have been included in their cover sets. For example, $\mathcal{N}_3=\{1,2,4\}$, and cover sets to which node $3$ belonging are $\{1,2,3\}$ and $\{3,4\}$. Obviously, $\mathcal{N}_3\subset\{1,2,3\}\cup\{3,4\}$. Similarly, Figure \ref{partition}b) is a star network system with $9$ nodes. One can find a cover as $\{1,2\}$, $\{1,3\}$, $\{1,4\}$, $\{1,5\}$, $\{1,6\}$, $\{1,7\}$, $\{1,8\}$, $\{1,9\}$ and all agents' physical neighbors are covered by their cover sets. As for Figure \ref{partition}c), the acquisition of cover sets is not so intuitive, but we can still obtain it with $\{1,2,3,4\}$, $\{5,6,7,8\}$, and $\{3,5\}$ after the observation and analysis. Through the above examples, we understand the basic idea and function of covering. However, for large-scale systems, we cannot obtain a reasonable coverage through mere observation. Therefore, a principled covering algorithm is necessary for distributed observers of large-scale systems.


To this end, we propose a coverage solving method based on the greedy algorithm. Before showing the algorithm details, some notations are introduced. $\mathcal{O}_k\subset\mathcal{V}$ is defined as the $k$th cover set of the large-scale system, and $\mathscr{P}_i$ denotes the set of cover sets that containing node $i$. For example, in Figure \ref{partition}c), $\mathcal{O}_1=\{1,2,3,4\}$, $\mathcal{O}_2=\{5,6,7,8\}$, and $\mathcal{O}_3=\{3,5\}$. In addition, $\mathscr{P}_1=\{\mathcal{O}_1\}$, and $\mathscr{P}_3=\{\mathcal{O}_1,\mathcal{O}_3\}$. Set $\mathcal{O}(\mathcal{P}_i)=\bigcup_{\mathcal{O}_j\in\mathscr{P}_i}\mathcal{O}_j$, and denote $D(\mathscr{P}_i)=\sum_{\mathcal{O}_j\in\mathscr{P}_i}|\mathcal{O}_j|$ the total number of agents that need to be estimated by local observer on the $i$th agent. Let $Pa(i,j)$ be a set of the nodes belonging to the shortest path between $i$ and $j$ in communication network $\forall i,j\in\mathcal{V}$. For example, in Figure \ref{partition}c), $Pa(1,4)=\{1,2,3,4\}$. {\color{blue}Additionally, we define for node $i$ indices $r_i$ and $s_i$, with the partial orders $r_i \preceq r_j$ if $|\mathcal{C}_{i}| < |\mathcal{C}_{j}|$ or ($|\mathcal{C}_{i}| = |\mathcal{C}_{j}|$ and $|\mathcal{N}_{i}| \geq |\mathcal{N}_{j}|$), and $s_i \succeq s_j$ if $|\mathcal{C}_{i}| > |\mathcal{C}_{j}|$ or ($|\mathcal{C}_{i}| = |\mathcal{C}_{j}|$ and $|\mathcal{N}_{i}| \geq |\mathcal{N}_{j}|$).}

\begin{algorithm}[!t]
	\caption{Coverage solving algorithm of distributed observer for large-scale systems} 
	\label{alg1}
	\begin{algorithmic}[1]
		\REQUIRE{$\mathcal{N}_i$ and $\mathcal{C}_i$ $\forall i\in\mathcal{V}$}
		\ENSURE{A cover set of distributed observer}
		\STATE \emph{Establishing cover sets based on greedy algorithm};
		\STATE Initialize $p=1$ and $\mathcal{O}_p=\emptyset$; initialize $\mathscr{P}_i=\emptyset$ $\forall i\in\mathcal{V}$; {\color{blue}initialize $\mathcal{V}_r=\mathcal{V}$ and $\mathcal{V}_r=\mathcal{V}$;}
		\WHILE{$|\mathcal{V}_r|\neq 0$}
		\STATE {\color{blue}Choose $i\in\mathcal{V}_r$ with minimum $r_i$;}
		\IF{$\mathcal{N}_{i}\backslash\mathcal{O}(\mathscr{P}_{r_i})\neq \emptyset$}
		\FOR{$j\in\mathcal{N}_{i}\backslash\mathcal{O}(\mathscr{P}_{r_i})$}
		\STATE Find $Pa(r_i,j)$;
		\STATE Set $\mathcal{O}_p=\mathcal{O}_p\cup Pa(r_i,j)$;
		\ENDFOR 
		\ENDIF
		\STATE Set $\mathscr{P}_{k}=\mathscr{P}_{k}\cup\{\mathcal{O}_p\}$ $\forall k\in\mathcal{O}_p$;
		\STATE Set $p=p+1$ and initialize $\mathcal{O}_p=\emptyset$;
		\STATE {\color{blue}Set $\mathcal{V}_r=\mathcal{V}_r\backslash\{i\}$;}
		\ENDWHILE
		\STATE \emph{Merging cover sets based on greedy algorithm};
		\WHILE{$|\mathcal{V}_s|\neq 0$}
		\STATE {\color{blue}Choose $i\in\mathcal{V}_s$ with maximum $s_i$;}
		\STATE Calculate a set $M_{i}$ and define $\ell_{i}=|M_{i}|$; 
		\STATE $//$Refer Algorithm 1 to obtain $M_{i}$
		\WHILE{$j\leq\ell_{i}$}
		\STATE Set the $j$th element of $M_{i}$ be $J_{ij}=\{p_{m1},p_{m2},\ldots,p_{m\ell}\}$, where $\ell=|J_{ij}|$;
		\IF{$\bigcup_{k=1}^\ell\mathcal{O}_{p_{mk}}=\emptyset$ or $\sum_{k=1}^\ell\min\{|\mathcal{O}_{p_{mk}}|,1\}=1$}
		\STATE Let $j=j+1$ and go to line $21$;
		\ENDIF
		\STATE Set $c_1=0$ and $c_2=0$;
		\STATE Calculate $D(\mathscr{P}_l)$ $\forall l\in\bigcup_{k=1}^\ell\mathcal{O}_{p_{mk}}$;
		\STATE Calculate $D^*(\mathscr{P}_{s_i})=\left|\bigcup_{k=1}^\ell\mathcal{O}_{p_{mk}}\right|$;
		\STATE Set $c_1=1$ if $D^*(\mathscr{P}_{s_i})-D(\mathscr{P}_l)\leq D(\mathscr{P}_{s_i})- D^*(\mathscr{P}_{s_i})$ $\forall l\in\bigcup_{k=1}^\ell\mathcal{O}_{p_{mk}}$ and $l\neq i$;
		\STATE Set $c_2=1$ if $\left(D^*(\mathscr{P}_{s_i})\right)^2\leq \sum_{l\in\bigcup_{k=1}^\ell\mathcal{O}_{p_{mk}}}D(\mathscr{P}_l)$;
		\IF{$c_1c_2=1$}
		\STATE Let $\mathcal{O}_{p_{m1}}=\bigcup_{k=1}^\ell\mathcal{O}_{p_{mk}}$, and $\mathcal{O}_{p_{mk}}=\emptyset$ for $k=2,\ldots,\ell$;
		\ENDIF
		\STATE Let $j=j+1$;
		\ENDWHILE
		\STATE {\color{blue}Set $\mathcal{V}_s=\mathcal{V}_s\backslash\{i\}$;}
		\ENDWHILE
	\end{algorithmic}
\end{algorithm}

Now, one can see Algorithm \ref{alg1} for the developed coverage solving method. To aid the reader's understanding, the following is a brief description of the algorithm.

\textbf{Step 1:}~Sort $r_i,~i=1,\ldots,N$ based on symbol $\preceq$.

\textbf{Step 2:}~{\color{blue}Each node traverses its physical network neighbors. If a neighbor is already in $\mathcal{O}(\mathscr{P}_i)$, no action is taken. If the neighbor is not in $\mathcal{O}(\mathscr{P}_i)$, all nodes on the shortest path from the current node to that neighbor are included in a new subset $\mathcal{O}_p$ (subsequently, any other newly added physical network neighbors and the nodes on their shortest paths will also be placed in this same $\mathcal{O}_p$; that is, each node creates only one new subset).}

\textbf{Step 3:}~Sort $s_i,~i=1,\ldots,N$ based on symbol $\succeq$.

\textbf{Step 4:}~Starting from the maximum $s_i$. Each node considers whether to merge several cover sets according to two criteria (Line 27--29 in Algorithm \ref{alg1} and their details are shown in (D2)) in sequence (In communication networks, nodes with larger degrees are more likely to be forced to join many redundant cover sets. Therefore, priority is given to nodes with larger degrees to determine whether to remove some redundant cover sets through merging).

Note that the coverage solving algorithms described in Algorithm \ref{alg1} and Algorithm \ref{alg2} actually solve a multi-objective optimization problem because each node has its own desired cover sets, but the desired optimal coverage among different nodes are conflicting. In the complete version of this article, we prove that the coverage obtained through Algorithm \ref{alg1} and Algorithm \ref{alg2} are Pareto solutions to this multi-objective optimization problem.

\begin{prop}
	Coverage obtained from Algorithm \ref{alg1} and Algorithm \ref{alg2} are already located on the Pareto frontier.
\end{prop}

\begin{IEEEproof}
	Given a group of cover sets ($\mathcal{O}_{1},\ldots,\mathcal{O}_{m}$) obtained by Algorithm 1 and Algorithm 2. We first prove three propositions.
	
	
	1)~Add any nodes in arbitrary $\mathcal{O}_i=\{i_1,\ldots,i_r\},~i\in\{1,2,\ldots,m\}$ would increase the dimensions of observers on at least one node. Since $\mathcal{O}_i$ is obtained by at least one node (denote as $i_1$) using a greedy algorithm, node $i_1$ makes the optimal choice when selecting $\mathcal{O}_i$ based on the current situation. Now, let's discuss two scenarios. First, $i_1$ has yet to be included in any cover set before selecting $\mathcal{O}_i$. In this case, $\mathcal{O}_i$ is the first cover set for $i_1$, so it is the optimal choice for $i_1$ at that time. In this scenario, adding any new node to this cover set will increase the dimension of the observer on $i_1$. Second, before selecting $\mathcal{O}_i$, $i_1$ has already been included in another cover set (denote as $\mathcal{O}_j$) by another node (let's call it $i_2$). In this case, if a new node $i_0$ is added into $\mathcal{O}_i$, then the dimension of observers on node $i_1$ will increase owing to the joining of $i_0$, unless $\mathcal{O}_j$ becomes a subset of $\mathcal{O}_i\cup \{i_0\}$ and node $i$ chooses to merge $\mathcal{O}_i$ and $\mathcal{O}_j$. However, $\mathcal{O}_j$ is the optimal choice made by $i_2$ based on the greedy algorithm. Therefore, any changes made by $i_1$ to $\mathcal{O}_j$ will inevitably result in an increase in the observer dimension on $i_2$. This proposition is proven.
	
	2)~For arbitrary $\mathcal{O}_i,~i\in\{1,2,\ldots,m\}$, every node within this set is indispensable. Taking $ i_\in\mathcal {O}_i$ as an example, there are two reasons why $i_1$ includes $i_2$ in the cover set based on the greedy algorithm. First, $i_1$ needs to estimate the information of $i_2$. In this case, if $i_2$ is excluded from the cover set, $i_1$ would still need to establish a new set $\mathcal{O}_i'$ that includes both $i_1$ and $i_2$ in order to estimate the state of $i_2$. Thus, it must hold that $|\mathcal{O}_{i}|<|\mathcal{O}_i'|+|\mathcal{O}_i\backslash\{i_2\}|=2+|\mathcal{O}_{i}|-1$. Second, if $i_1$ does not need to estimate the state of $i_2$, it implies that the connectivity of the communication network within $\mathcal{O}_i$ depends on the presence of $i_2$. In this case, if $i_2$ is removed, the cover set becomes invalid. In conclusion, we have demonstrated that ``any node in any cover set is indispensable".
	

	Based on the above three propositions, we will prove that the coverage obtained by Algorithm \ref{alg1} and Algorithm \ref{alg2} are already Pareto optimal solutions. To this end, it is sufficient to show that reducing the value of $D(\mathscr{P}_j)$ for any node (e.g., $j$) will result in an increase in $D(\mathscr{P}_k)$ for at least one node (e.g., $k$). We will discuss this with two cases.
	
	First, let's assume that $\mathscr{P}_j$ contains multiple cover sets ($\mathcal{O}_{j1}$, $\mathcal{O}_{j2}$, $\ldots$). In order to reduce $D(\mathscr{P}_{j})$, node $j$ can either remove a node from a certain set (e.g., deleting node $k$ from $\mathcal{O}_{j1}$) or merge multiple cover sets (e.g., merging $\mathcal{O}_{j1}$ and $\mathcal{O}_{j2}$). Based on the previous propositions, there exists at least one node $i_1$, for which $\mathcal{O}_{j1}$ is optimal. Therefore, any operation of reducing or merging $\mathcal{O}_{j1}$ will inevitably lead to an increase in $D(\mathscr{P}_{i_1})$. Hence, in this case, optimizing the observer dimensions of node $j$ will inevitably result in at least one other node suffering a loss in benefit or resulting in an invalid coverage.
	
	Second, let's assume that $\mathscr{P}_j$ contains only one cover set. In this case, the only way for node $j$ to reduce $D(\mathscr{P}_{j})$ is by removing some nodes from that cover set. However, according to Proposition 2), this approach will obviously result in at least one other node suffering a loss in benefit.
	
	In conclusion, the coverage obtained from Algorithm \ref{alg1} and Algorithm \ref{alg2} are already located on the Pareto frontier.
\end{IEEEproof}

We have the following discussions regarding Algorithm \ref{alg1} and \ref{alg2}.

\textbf{(D1)}~The node constraint conditions can be satisfied. The guarantee of constraint conditions (It is required that the union of all cover sets where each agent is located should include all its physical network neighbor agents) is mainly in the first part ``establishing cover sets" Algorithm \ref{alg1} because its ``cover sets initialization" serves the purpose of allowing each node to select sets to cover its physical network neighboring nodes.

\begin{algorithm}[!t]
	\caption{Calculate $M_{s_i}$ for $i=1,\ldots,N$} 
	\label{alg2}
	\begin{algorithmic}[1]
		\REQUIRE{$\mathscr{P}_{i}$}
		\ENSURE{$M_{i}$}
		\STATE Initialize $M_{i}=\emptyset$;
		\STATE Construct a set of $\mathscr{P}_{i}$'s subsets, denoted by $\mathscr{P}_{i}^c=\{\mathcal{J}\subset\mathscr{P}_{i};~|\mathcal{J}|\geq 2\}$; Without lossing of generality, define $\mathscr{P}_{i}^c=\{\mathcal{J}_{i1},\mathcal{J}_{i2},\ldots,\mathcal{J}_{i\zeta_{i}}\}$, where $\zeta_{i}=2^{|\mathscr{P}_{i}|}-|\mathscr{P}_{i}|-1$;
		\FOR{$k=1$ to $\zeta_{i}$}
		\STATE Calculate $\mathcal{I}(\mathcal{J}_{ik})=\bigcap_{\mathcal{O}_p\in\mathcal{J}_{ik}}\mathcal{O}_p$;
		\IF{$|\mathcal{I}(\mathcal{J}_{ik})|\geq 2$}
		\STATE Set $M_{i}=M_{i}\cup\{\mathcal{J}_{ik}\}$;
		\ENDIF
		\ENDFOR
	\end{algorithmic}
\end{algorithm}

\textbf{(D2)}~The central part of Algorithm \ref{alg1} adopts the idea of a greedy algorithm. According to (D1), the feasible coverage under the constraint conditions has been reached with the first part of Algorithm \ref{alg1}. However, since the first part starts from the node with the smallest degree in the communication network, the node with a large degree (hub node) may have an excessive computational burden. The second part of the algorithm is mainly employed to reduce the computational burden of hub nodes. The developed method allows the hub nodes' neighbors in the communication network to share the computing burden equally by merging the cover sets of hub nodes. Whether it is shared equally depends on: 

1) The increment of the local observer dimension of any neighbor node cannot be greater than the reduction of the local observer dimension of the hub node; 

2) The total dimension of the local observers of the hub node and its neighbors should be reduced compared to that before merging cover sets. 

Therefore, the merging part will further optimize the coverage obtained by cover sets initialization.

\textbf{(D3)}~Description of the ``Merging Cover Set" part. Suppose a hub node wants to merge cover sets $\mathcal{O}_{p_{m1}},\ldots,\mathcal{O}_{p_{m\ell}}$. After the sets are merged, the local observers of all nodes in the merged set $\mathcal{O}_{p_{m1}}$ must estimate all the states involved in $\mathcal{O}_{p_{m1}}$. Hence, the dimensions of local observers of all nodes are equal after merging, which can be denoted as $D^*(\mathscr{P}_{s_i})=|\bigcup_{k=1}^\ell\mathcal{O}_{p_{mk}}|$ where $s_i$ is the index of the hub node. Therefore, condition 1) in (D2) indicates $D^*(\mathscr{P}_{s_i})-D(\mathscr{P}_l)\leq D(\mathscr{P}_{s_i})- D^*(\mathscr{P}_{s_i})$ for $l=1,\ldots,\ell_{s_i}$. Furthermore, after merging cover sets, there are $D^*(\mathscr{P}_{s_i})$ nodes in the set, and each node needs to estimate all states of $D^*(\mathscr{P}_{s_i})$ nodes. Therefore, condition 2) can be formulated as $\left(D^*(\mathscr{P}_{s_i})\right)^2\leq\sum_{l\in\bigcup_{k=1}^\ell\mathcal{O}_{p_{mk}}}D(\mathscr{P}_l)$. The aforementioned explains the source of step $27$ and step $28$ in Algorithm \ref{alg1}.

\begin{figure*}[!t]
	\centering
	\includegraphics[width=17cm]{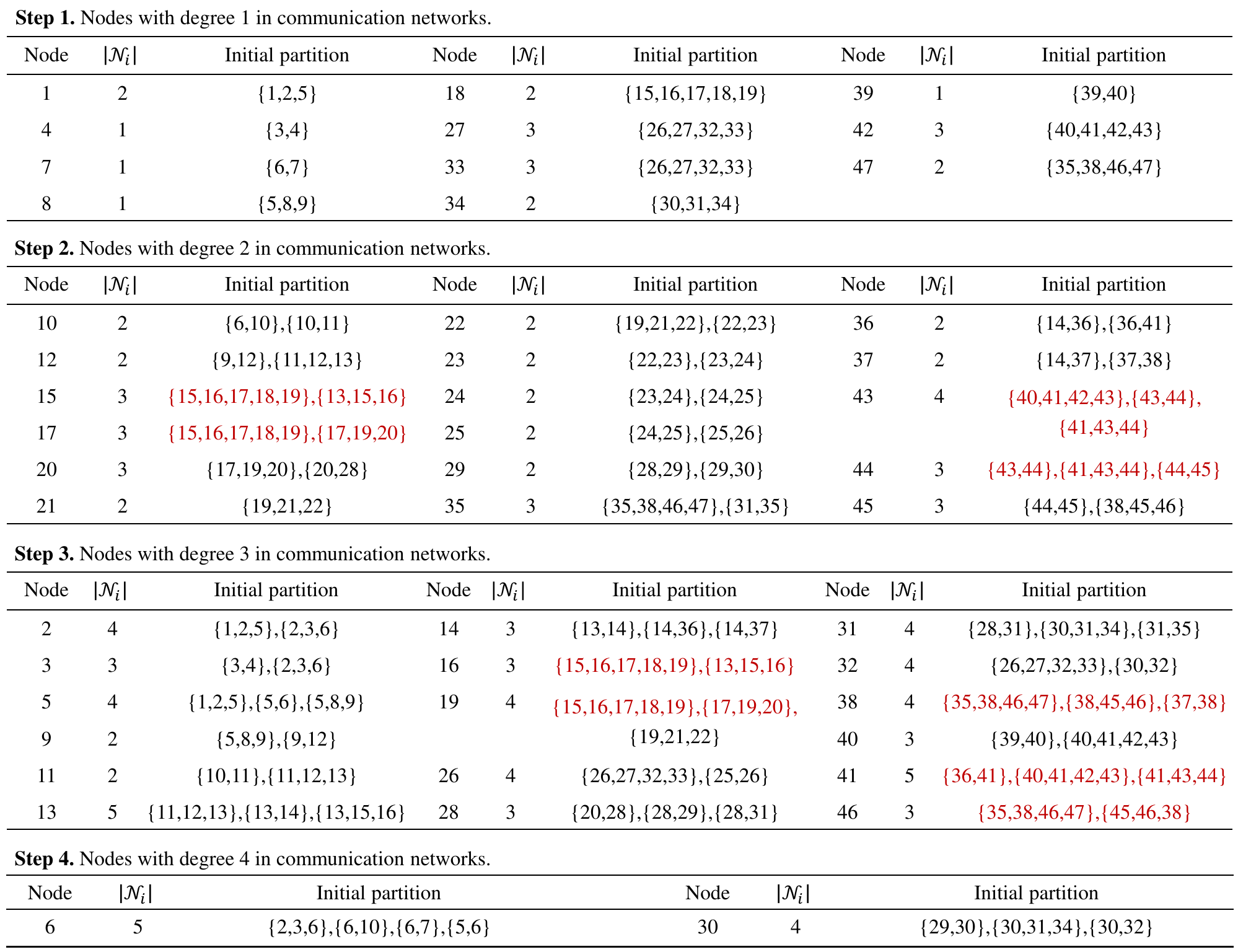}\\
	\caption{Usage example of Algorithm \ref{alg1}: initial cover set. The part marked in red is the node with $\ell_i\neq 0$, which needs to consider whether to merge covers sets.}\label{example1}
\end{figure*}
\begin{figure*}[!t]
	\centering
	\includegraphics[width=17cm]{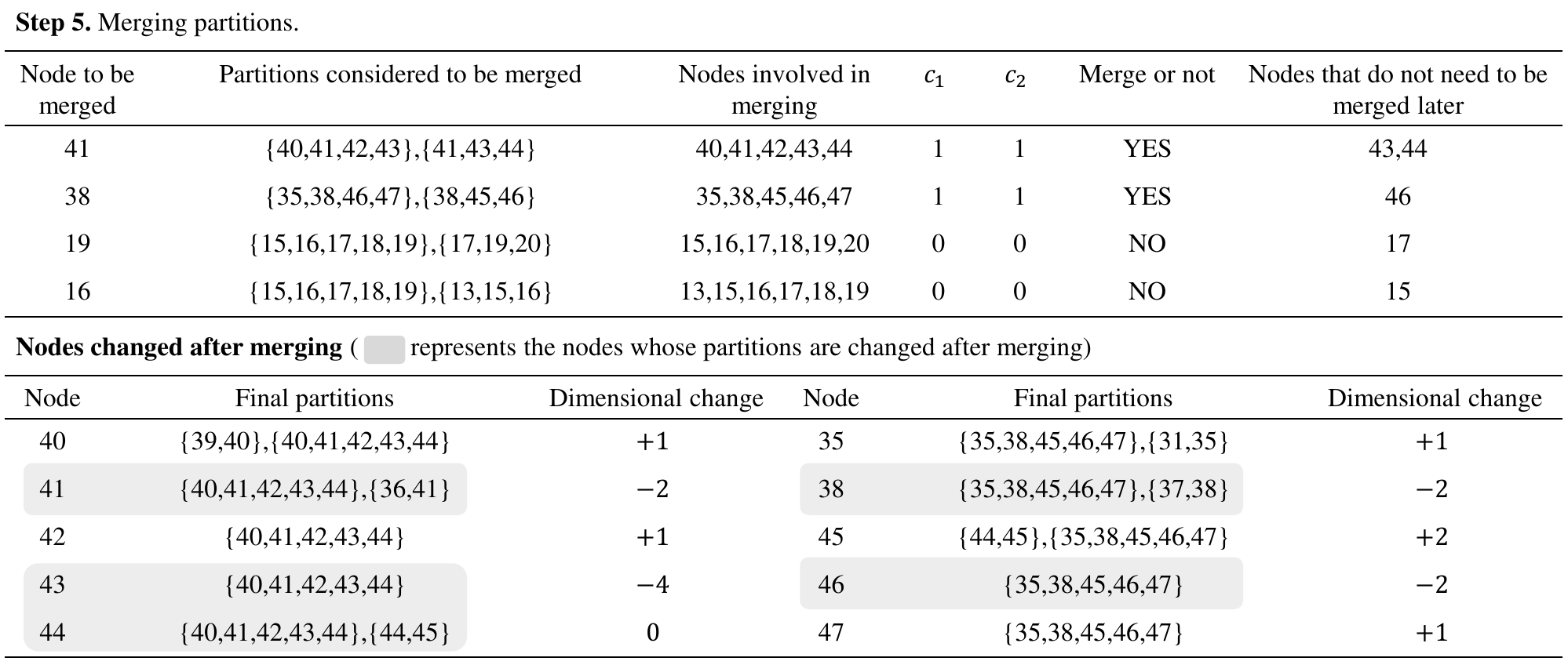}\\
	\caption{Usage example of Algorithm \ref{alg1}: Final coverage.}\label{example2}
\end{figure*}

\textbf{(D4)}~Algorithm \ref{alg1} has polynomial time complexity. 

First, focus on the first part of the algorithm. The complexity of this part is reflected in the number of calculations of lines $7$ to line $8$. They are the core contents of ``establishing coverage" and are calculated with at most $N\cdot\frac{1}{N}\sum_{i=1}^N|\mathcal{N}_{i}|$ times. Note that the average degree $\frac{1}{N}\sum_{i=1}^N|\mathcal{N}_{i}|$ is scarce relative to the total number $N$ in a large-scale network. 

Second, we calculate the time complexity of Algorithm \ref{alg2}. Since Algorithm \ref{alg2} is the $19$th line of Algorithm \ref{alg1}, its complexity is also a part of the complexity of Algorithm \ref{alg1}. Algorithm \ref{alg2}'s core steps include line $4$ and line $6$, and they are calculated with at least $N\cdot \frac{1}{N}\sum_{i=1}^N\zeta_i$ times. According to the coverage solving method, the number of cover sets of each node will not exceed its communication network degree. Hence, 
\begin{align}
	\frac{1}{N}\sum_{i=1}^N\zeta_i=&\frac{1}{N}\sum_{i=1}^N\left(2^{|\mathscr{P}_i|}-|\mathscr{P}_i|-1\right)\notag\\
	\approx&\frac{1}{N}\sum_{i=1}^N\left(2^{|\mathcal{C}_i|}-|\mathcal{C}_i|-1\right)\notag\\
	=&2^{\kappa N}-\kappa N-1,
\end{align}
where $\kappa$ stands for the proportion of the average degree of communication network to $N$. Note that $2^{\kappa N}-\kappa N-1=O(N)$ when $\kappa\leq \frac{1}{N}\log_2[N+1]\ll N$.

Third, core steps (line $26,27,31$) of the last loop (line $16$ to line $36$) of Algorithm \ref{alg1} should be calculated with $N\cdot\frac{1}{N}\sum_{i=1}^N\ell_{s_i}$ times. Additionally,
\begin{align}
	\frac{1}{N}\sum_{i=1}^N\ell_{i}<\frac{1}{N}\sum_{i=1}^N\zeta_i=O(N).
\end{align}

In summary, the complexity of Algorithm \ref{alg1} is
\begin{align}
	&O(N)\frac{1}{N}\sum_{i=1}^N|\mathcal{N}_{i}|+O(N)\frac{1}{N}\sum_{i=1}^N\zeta_i+O(N)\frac{1}{N}\sum_{i=1}^N\ell_{i}\notag\\
	\leq&O(N)+O(N)\cdot O(N)+O(N)\cdot O(N)\leq O(N^2).
\end{align}
Therefore, Algorithm \ref{alg1} has the polynomial time complexity.

The following text explains Algorithm \ref{alg1} and \ref{alg2} in more detail with an example. Figure \ref{example1} shows the physical and communication network considered in this example, and  The detailed steps of find coverage are listed in Figure \ref{example1} and Figure \ref{example2}, where the former shows the initial cover sets steps and the latter shows the merging steps. To facilitate readers' understanding, some key steps will be interpreted.

\textbf{(I1)}~We focus on the area containing nodes $1,\ldots,10$. This area is taken as an example to show why cover set selecting should be carried out in the order of $r_1\preceq r_2\preceq\cdots\preceq r_N$. In light of the sorting rule, node $1$ with $\mathcal{N}_1=2$ and $\mathcal{C}_1=1$ is the first node to be considered. Based on line $7$ to line $10$, we know $\mathcal{O}_1=\{1,2,5\}$. Hence, $\mathscr{P}_1=\{\mathcal{O}_1\}$, $\mathscr{P}_2=\{\mathcal{O}_1\}$, $\mathscr{P}_5=\{\mathcal{O}_1\}$. Subsequently, sets $\mathcal{O}_2=\{3,4\}$, $\mathcal{O}_3=\{6,7\}$, $\mathcal{O}_4=\{5,8,9\}$, $\mathcal{O}_5=\{6,10\}$, $\mathcal{O}_6=\{10,11\}$ can be selected for nodes $4$, $7$, $8$, and $10$ in order. Then, note that $1,5\in\mathcal{O}(\mathscr{P}_2)$; thus $\mathcal{O}_7=\{2,3,6\}$ is necessary included in $\mathscr{P}_2$. Besides, since $\mathcal{N}_3$ has been contained in $\mathcal{O}(\mathscr{P}_3)$, node $3$ needs not to select sets any more. This reflects the advantages of the proposed node sorting rule: the node with a larger degree in the communication network selects the set in the later order, which neither affects its cover set selection nor interferes with the cover set selection of the previous node, but greatly reduces the computational load (most of the required cover sets have been selected in advance). The same situation can also be seen at node $6$. However, the unreasonable order of nodes will lead to a large number of cover sets with high repetition (See in I2).

\textbf{(I2)}~We focus on nodes $42$, $43$, and $44$. Their selection order is $42\preceq 43\preceq 44$. Node $42$ selects its cover sets as $\{40,41,42,43\}$. Hence, only node $44$ remains in $\mathcal{N}_{43}\backslash\mathcal{O}(\mathscr{P}_{43})$. It means $\{43,44\}$ should be added into $\mathscr{P}_{43}$. However, since $41\in\mathcal{N}_{44}$, $\mathscr{P}_{44}$ must include a set $\{41,43,44\}$. It leads to a new set to be added to node $43$, which has completed the cover set selection before that. More unreasonably, two highly repetition cover sets $\{41,43,44\},\{43,44\}$ exist at the same time in $\mathscr{P}_{43}$ and $\mathscr{P}_{44}$. The reason for this phenomenon is that $|\mathcal{N}_{44}\backslash\mathcal{O}(\mathscr{P}_{44})|>|\mathcal{N}_{43}\backslash\mathcal{O}(\mathscr{P}_{43})|$ after the node $42$ selects its cover sets. The above analysis shows that we have the advantage of allowing the node with a smaller communication network degree and larger physical network degree to select cover sets preferentially. Otherwise, highly similar cover sets like $\{41,43,44\}$ and $\{43,44\}$ will appear in large numbers. In addition, the above analysis also shows that a more reasonable sorting rule should be: node $i$ will preferentially select the cover set if $|\mathcal{N}_{i}\backslash\mathcal{O}(\mathscr{P}_{i})|>|\mathcal{N}_{j}\backslash\mathcal{O}(\mathscr{P}_{j})|$ when $|\mathcal{C}_j|=|\mathcal{C}_i|$. However, this sort rule means that one has to sort the nodes again after a node selects its cover sets, which is adverse to the implementation of the algorithm. Therefore, we do not improve the algorithm from the perspective of the remaining physical network neighbor nodes but use the merging strategy to make up for this problem.

\textbf{(I3)}~Figure \ref{example2} displays the steps of merging cover sets. Note that the actual calculation is not complex, although the complexity of the merging part is $O(N^2)$. It is because after the cover sets of some nodes are merged, many other nodes that initially needed to merge cover sets no longer need to perform the merging steps. For example, in Figure \ref{example1}, all nodes $41,43,44$ need to merge their cover sets. However, $M_{s_i}$ (defined by Algorithm \ref{alg2}) with respect to node $43$ and $44$ are both empty after $41$ merging its sets, and thus they need not perform the merging steps anymore.

\begin{rem}
	From this example, we can see that if we design cover-based distributed observers based on the coverage shown in Figure \ref{example1} and \ref{example2}, we can significantly reduce the dimension of observers on each agent. However, we must emphasize an important fact: the degree of reduction in observer dimension on agents is closely related to the similarity between the communication network and the physical network. Due to the arbitrariness of communication and physical networks, some strange coverage may occur. For example, two nodes that are neighbors in a physical network may be far apart in a communication network, which can lead to the occurrence of huge cover set. In such cases, the reduction in the dimension of observers on each agent will not be as pronounced as in Figure \ref{example1} and \ref{example2}. Therefore, we introduce the concept of ``network similarity"---defined by $S_{pc}=2|\mathcal{E}_c\cap\mathcal{E}_p|/(|\mathcal{E}_c|+|\mathcal{E}_p|)$---to improve the rigor when describing simulation results. In other words, when describing how much our algorithm reduces the dimension of observers on each agent, we must indicate the degree of network similarity because the lower the network similarity, the higher the possibility of large abnormal cover sets, and the lower the possibility of significant reduction in observer dimension.
\end{rem}

\subsection{Cover-based distributed observer}\label{sec3.2}

This subsection designs distributed observer for each cover set based on the coverage solving method in the previous subsection. Suppose that $\mathcal{O}_p$ is a cover set belonging to $\mathscr{P}_i$, and then the state observer on the $i$th agent to itself takes the form of:
\begin{align}
	\dot{\hat{x}}_{ii}^{(p)}=&A_{ii}\hat{x}_{ii}^{(p)}+\sum_{j\in\mathcal{N}_i}A_{ij}\bar{x}_{ij}+\theta\Gamma_\theta^{-1} H_i\left(y_i-C_i\hat{x}_{ii}^{(p)}\right)\notag\\
	&+B_i\hat{u}_{ii}+\gamma\theta^{n-1}\tilde{P}_i^{-1}\sum_{j\in\mathcal{O}_p}\alpha_{ij}\left(\hat{x}_{ji}^{(p)}-\hat{x}_{ii}^{(p)}\right),\label{observer1}
\end{align}
where $\hat{x}_{ii}^{(p)}$ is the estimation of $x_i$ generated by observers on agent $i$ in cover set $\mathcal{O}_p$; $\bar{x}_{ij}=1/N_{j,\mathscr{P}_i}\cdot\sum_{p,~\mathcal{O}_p\in\mathscr{P}_i}\hat{x}_{ij}^{(p)}$ with $N_{j,\mathscr{P}_i}$ being the number of occurrences of $j$ in $\mathcal{O}(\mathscr{P}_i)$; $\hat{x}_{ij}^{(p)}$ stands for the state estimation of the $j$th subsystem generated by the $i$th agent. $\Gamma_\theta=diag\{\theta^{n-1},\ldots,\theta,1\}$ is the high-gain matrix with $\theta$ being the high-gain parameter; $H_i\in\mathbb{R}^{n\times p}$ and $\gamma$ are, respectively, the observer gain and coupling gain; $\tilde{P}_i=\Gamma_\epsilon P_i\Gamma_\epsilon^{-1}$ is the so-called weighted matrix and $P_i$ is a symmetric positive definite matrix that will be designed later; $\hat{u}_{ii}^{(p)}$ is the control input relying on $\bar{x}_{ij}$. 

The dynamics of $\hat{x}_{li}^{(p)}$ $\forall l\in\mathcal{O}_p$ can be expressed as
\begin{align}
	\dot{\hat{x}}_{li}^{(p)}=&A_{ii}\hat{x}_{li}^{(p)}+\sum_{j\in\mathcal{O}(\mathscr{P}_l)\cap\mathcal{N}_i}A_{ij}\bar{x}_{lj}\notag\\
	&+B_i\hat{u}_{li}+\gamma\theta^{n-1}\sum_{j\in\mathcal{O}_p}\alpha_{ij}\left(\hat{x}_{ji}^{(p)}-\hat{x}_{li}^{(p)}\right),\label{observer2}
\end{align}
where $\hat{x}_{li}^{(p)}$ is the estimation of $x_i$ generated by observers on agent $l$ in cover set $\mathcal{O}_p$, and $\hat{u}_{li}$ is the control input relying on $\bar{x}_{lj}$. 

Denote $e_{ii}^{(p)}=x_i-\hat{x}_{ii}^{(p)}$ and $e_{li}^{(p)}=\hat{x}_{li}^{(p)}-x_i$, we have
\begin{align}
	\dot{e}_{ii}^{(p)}=&\left(A_{ii}-\theta\Gamma_\theta^{-1} H_iC_i\right)e_{ii}^{(p)}+\sum_{j\in\mathcal{N}_i}A_{ij}\bar{e}_{ij}\notag\\
	+&B_i(\hat{u}_{ii}-\bar{u}_i)-\gamma\theta^{n-1} \tilde{P}_i^{-1}\sum_{j\in\mathcal{O}_p}\alpha_{ij}\left(e_{ji}^{(p)}-e_{ii}^{(p)}\right),\label{e1}\\
	\dot{e}_{li}^{(p)}=&A_{ii}e_{li}^{(p)}+\sum_{j\in\mathcal{O}(\mathscr{P}_l)\cap\mathcal{N}_i}A_{ij}\bar{e}_{lj}\notag\\
	+&B_i(\hat{u}_{li}-\bar{u}_i)-\gamma\theta^{n-1}\sum_{j\in\mathcal{O}_p}\alpha_{ij}\left(e_{ji}^{(p)}-e_{li}^{(p)}\right),\label{e2}
\end{align}
where $\bar{e}_{lj}=1/N_{j,\mathscr{P}_i}\cdot\sum_{\mathcal{O}_p\in\mathscr{P}_l}e_{lj}^{(p)}$; and $\bar{u}_i$ is the actual control law employed in the closed-loop system.

\begin{rem}
	This section has achieved the first goal of this paper through designing the coverage solving algorithm and cover-based distributed observer: the dimension of observers on agent $i$ ($\hat{x}_{i\star}\triangleq col\{\hat{x}_{ij}^{(p)},~j\in\mathcal{O}_p,~\mathcal{O}_p\in\mathscr{P}_i\}$) is much smaller than that of the interconnected system. 
\end{rem} 
\begin{rem}\label{r3}
	There are two differences between the design of the cover-based distributed observer and the traditional distributed observer \cite{2019Completely,Han2017A}. First, since the states of the $j$th subsystem may be repeatedly estimated by agent $i$ in multiple cover sets, we introduce the fusion estimation ($\bar{x}_{ij}$) in the coupling part of  observers on each agent. Second, agent $i$ needs to use the physical network neighbor information of agent $l$ when estimating the states of subsystem $l$ (see in (\ref{observer2})). In traditional distributed observers, this information can be directly found in observers on the $i$th agent. However, since the dimension of observers on the $i$th agent in this paper is much smaller than that of the interconnected system, its states fail to cover the neighbor states of subsystem $l$. Therefore, we need to eliminate the coupling relationships ($\sum_{j\in\mathcal{O}(\mathscr{P}_l)\cap\mathcal{N}_i}A_{ij}\bar{x}_{lj}$ in (\ref{observer2})) that cannot be obtained by agent $i$ when designing $\hat{x}_{il}$. In other words, to reduce the dimension of observers on each agent, we have to artificially introduce model mismatch to compensate for the lack of information caused by observer dimension reduction. Moreover, the control input $\hat{u}_{il}$ in observers on the $i$ agent also needs the state estimation $\hat{x}_{l\star}$. So, multiple estimates and model mismatch also occur in the control input ($\hat{u}_{ii}$ and $\hat{u}_{li}$) of the cover-based distributed observer (\ref{observer1}) and (\ref{observer2}). The content introduced here is the novelty of the cover-based distributed observer design in this section, and it is also one of the main difficulties to be solved later.
\end{rem}

\section{Distributed control law for large-scale linear systems}\label{sec4}
Section \ref{sec3} has reduced the dimension of observers on each agent by find a coverage of all agents. This section will analyze the performance of the cover-based distributed observer and the closed-loop system after the dimension reduction of the local observer. Subsection \ref{sec4.1} shows the design of distributed control law. The stability of the error dynamics (\ref{e1}) and (\ref{e2}) and the closed-loop system will be proved in subsection \ref{sec4.2} and \ref{sec4.3}, respectively. Subsection \ref{sec4.4} will show the achievement of the second goal of this paper.

\subsection{Distributed control law}\label{sec4.1}
{\color{cyan}A state feedback control law can be designed for the large-scale system described by (\ref{sys21}) and (\ref{sys22}). Considering the dynamics of the $i$-th subsystem in (\ref{sys11}) and (\ref{sys12}), a globally asymptotically stable control law takes the form:
	\begin{align}\label{u1}
		u_i=\sum_{j\in\mathcal{N}_i\cup\{i\}}K_{ij}x_j,
	\end{align}
where the gain matrices $K_{ij}$ are designed using the LQR method, with $K=[K_{ij}]_{i,j=1}^N$ representing the complete centralized optimal gain matrix. For network systems where the dynamics and cost matrices exhibit a Spatially-Exponential Decaying (SED) structure \cite{zhang2025optimal}, the optimal gain matrix $K$ possesses a quasi-SED property (or Quasi-Sparsity). Specifically, the norm of the coupling gain $\|K_{ij}\|$ decays exponentially with the distance between subsystem $i$ and $j$, $\text{dist}(i,j)$. Thus, the terms for $j\notin \mathcal{N}_i$ are indeed sufficiently small. More importantly, theoretical analyses, such as those in \cite{zhang2025optimal}, demonstrate that when employing a truncated local controller as (\ref{u1}) (only use physical neighbor's states), the performance gap between this distributed control law and the global optimal centralized control is exponentially small in $2$. Therefore, (\ref{u1}) can be regarded as an approximate centralized control law.
}

However, (\ref{u1}) is the state feedback control law with precise system states. Hence, one should replace them with the state estimations generated by the observers on the $i$th agent. In the case of the traditional distributed-observer-based distributed control law \cite{Xu2022TIV,Xu2020Distributed,8985536,2021DistributedMeng}, the exact states of (\ref{u1}) can be directly replaced by state estimations of the $i$th local observer. However, the concept of coverage is introduced and thus leads to a fusion estimation problem (See Remark \ref{r3}. For example, agent $31$ in the example of Figure \ref{example1} and Figure \ref{example2} has estimated its states three times in three cover sets). Hence,  $\bar{x}_{ii}=1/|\mathscr{P}_i|\cdot\sum_{\mathcal{O}_p\in\mathscr{P}_i}\hat{x}_{ii}^{(p)}$, and $\bar{x}_{li}=1/N_{i,\mathscr{P}_l}\cdot\sum_{\mathcal{O}_p\in\mathscr{P}_l}\hat{x}_{li}^{(p)}$ (note that $1/N_{i,\mathscr{P}_i}=1/|\mathscr{P}_i|$) are expected to be employed in the control law of the $i$th subsystem, i.e.,
\begin{align}\label{u2}
	\hat{u}_{ii}=\sum_{j\in\mathcal{N}_i\cup\{i\}}K_{ij}\bar{x}_{ij}.
\end{align}
Control law (\ref{u2}) can be used in observers on the $i$th agent but not in closed-loop systems because it cannot protect the closed-loop performance from the peak phenomenon of observers. Hence, we need to add a saturation mechanism, i.e.,
\begin{align}\label{u3}
	\bar{u}_{i}=\sum_{j\in\mathcal{N}_i\cup\{i\}}K_{ij}\mathbbm{x}_{ij},
\end{align}
where $\mathbbm{x}_{ij}$ is the so-called saturation value of $\bar{x}_{ij}$ and it is
expressed as $\mathbbm{x}_{ij}=\mathcal{M}sat\{\bar{x}_{ij}/\mathcal{M}\}$ with $\mathcal{M}$ being a given positive
constant and $sat\{\cdot\}$ being the saturation function. Furthermore, $\hat{u}_{li}$ in (\ref{observer2})---the control law with the state estimations of the $l$th agent to the $i$th subsystem---takes the form of:
\begin{align}\label{u4}
	\hat{u}_{li}=\sum_{j\in\mathcal{O}(\mathscr{P}_l)\cap\mathcal{N}_i}K_{ij}\bar{x}_{lj}.
\end{align}

Control law (\ref{u1}) is with precise and global states and is thus regarded as the quasi  centralized control law, which has the almost same performance of centralized control law. (\ref{u2}) and (\ref{u4}) are employed in distributed observer (\ref{observer1}) and (\ref{observer2}). As for (\ref{u3}), it is the distributed-observer-based distributed control. The closed-loop system combined with (\ref{sys21}) and (\ref{u3}) is the main research object of this paper. 
\begin{figure}[!t]
	\centering
	\includegraphics[width=8.5cm]{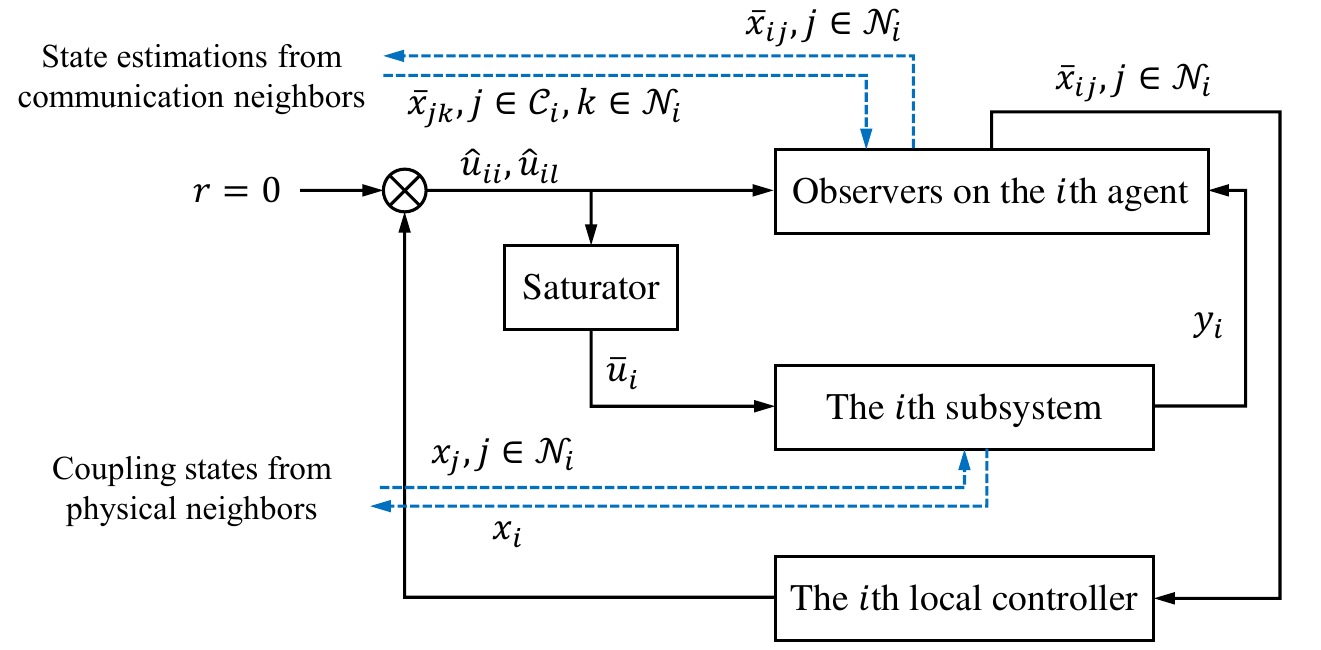}\\	
	\caption{Schematic diagram for delineating the	observation and control strategies.}\label{diagram}
\end{figure}

For the convenience of understanding, the diagram of the distributed control law described in this subsection is shown in Figure \ref{diagram}. As illustrated, the observers on the $i$th agent use $y_i$ and the information obtained via the communication network to obtain $\bar{x}_{ij},~j\in\mathcal{N}_i$ required for the $i$th agent. Then, observers send these $\bar{x}_{ij}$ to the $i$th local controller to get $\hat{u}_{ii}$ and $\hat{u}_{il},~l\in\mathcal{O}(\mathscr{P}_i)$. Then, these control signals are used in observers of $i$th agent, and $\bar{u}_i$---control signal after applying saturation mechanism---is used to control the $i$th subsystem.

In what follows, some properties concerning these four control laws (\ref{u1})--(\ref{u4}) are proved.

\begin{lem}\label{lem1}
	There exists constants $T_1>0$, $\rho_{ii}>0$, and $\kappa_{ii}>0$ such that
	\begin{align}
		\|\hat{u}_{ii}-\bar{u}_i\|\leq\rho_{ii}\|e_{i\star}\|+\kappa_{ii},
	\end{align}
	for $t<T_1$ and all $i=1,\ldots,N$, where $e_{i\star}=col\{e_{ij}^{(p)},~j\in\mathcal{O}_p,~\mathcal{O}_p\in\mathscr{P}_i\}$.
\end{lem}
\begin{IEEEproof}
	Since there is no finite time escape problem for linear systems, a constant $T_1>0$ exists such that the states of closed-loop system (\ref{sys21}) and (\ref{u3}) are bounded when $t<T_1$. Then, we know there are constants $M_{u_i}>0$ and $M_{\bar{u}_i}$ such that $\|u_i\|\leq M_{u_i}$ and $\|\bar{u}_i\|\leq M_{\bar{u}_i}$ when $t<T_1$. It follows $\|u_i-\bar{u}_{i}\|\leq M_{\bar{u}_i}+M_{u_i}$. To continue, we first calculate
	\begin{align*}
		&\left\|col\left\{\bar{e}_{ij},~j\in\mathcal{N}_i\cup\{i\}\right\}\right\|^2\notag\\
		\leq&\sum_{j\in\mathcal{N}_i\cup\{i\}}\left\|\sum_{\mathcal{O}_p\in\mathscr{P}_i}e_{ij}^{(p)}\right\|^2\leq 2\sum_{j\in\mathcal{N}_i\cup\{i\}}\sum_{\mathcal{O}_p\in\mathscr{P}_i}\left\|e_{ij}^{(p)}\right\|^2\notag\\
		\leq&\sum_{\mathcal{O}_p\in\mathscr{P}_i}\sum_{j\in\mathcal{O}_p}\left\|e_{ij}^{(p)}\right\|^2\leq 2\|e_{i\star}\|^2.
	\end{align*}
	Consequently,
	\begin{align}\label{uii-ui}
		&\|\hat{u}_{ii}-u_i\|\notag\\
		=&\left\|\sum_{j\in\mathcal{N}_i\cup\{i\}}K_{ij}\bar{x}_{ij}-\sum_{j\in\mathcal{N}_i\cup\{i\}}K_{ij}x_j\right\|\notag\\
		=&\left\|\sum_{j\in\mathcal{N}_i\cup\{i\}}K_{ij}\frac{1}{N_{j,\mathscr{P}_i}}\sum_{\mathcal{O}_p\in\mathscr{P}_i}\hat{x}_{ij}^{(p)}-\sum_{j\in\mathcal{N}_i\cup\{i\}}K_{ij}x_j\right\|\notag\\
		=&\left\|\sum_{j\in\mathcal{N}_i\cup\{i\}}K_{ij}\frac{1}{N_{j,\mathscr{P}_i}}\sum_{\mathcal{O}_p\in\mathscr{P}_i}e_{ij}^{(p)}\right\|\notag\\
		\leq&\left\|[K_{ij},~j\in\mathcal{N}_i\cup\{i\}]\right\|\times\left\|col\left\{\bar{e}_{ij},~j\in\mathcal{N}_i\cup\{i\}\right\}\right\|\notag\\
		\leq&\sqrt{2}\|K_{i}\|\|e_{i\star}\|,
	\end{align}
	where $K_i=[K_{i1},\ldots,K_{iN}]$. Denote $\rho_{ii}=\sqrt{2}\|K_{i}\|$, and $\kappa_{ii}=M_{\bar{u}_i}+M_{u_i}$. Then, we obtain
	\begin{align*}
		\|\hat{u}_{ii}-\bar{u}\|=\|\hat{u}_{ii}-u_i\|+\|u_i-\bar{u}_{i}\|
		\leq\rho_{ii,2}\|e_{i\star}\|+\kappa_{ii}.
	\end{align*}
	This completes the proof. 
\end{IEEEproof}

\begin{lem}\label{lem2}
	There are constants $T_1>0$, $\rho_{li}>0$, $\kappa_{ii}>0$ such that
	\begin{align}
		\|\hat{u}_{li}-\bar{u}_i\|\leq\rho_{li}\|e_{l\star}\|+\kappa_{li},
	\end{align}
	for $t<T_1$ and all $l\in\mathcal{O}_p$ and $\mathcal{O}_p\in\mathscr{P}_i$, where  $e_{l\star}=col\{e_{lj}^{(p)},~j\in\mathcal{O}_p,~\mathcal{O}_p\in\mathscr{P}_l\}$.
\end{lem}
\begin{IEEEproof}
	Similar to the proof of Lemma \ref{lem1}, there are $T_1>0$ and $W_{T_1}>0$ such that $x_i$ $\forall i=1,\ldots,N$ are bounded by $W_{T_1}$ when $t<T_1$. Accordingly, we directly calculate
	\begin{align}
		&\|\hat{u}_{li}-u_i\|\notag\\
		\leq&\left\|\sum_{j\in\mathcal{N}_i\cap\mathcal{O}(\mathscr{P}_l)}K_{ij}\frac{1}{N_{j,\mathscr{P}_l}}\sum_{\mathcal{O}_p\in\mathscr{P}_l}\hat{x}_{lj}^{(p)}\right.\notag\\
		&\left.-\sum_{j\in\mathcal{N}_i\cap\mathcal{O}(\mathscr{P}_l)}K_{ij}x_j-\sum_{j\in\mathcal{O}(\mathscr{P}_l)\backslash(\mathcal{N}_i\cup\{i\})}K_{ij}x_j\right\|\notag\\
		\leq&\sqrt{2}\|K_i\|\|e_{l\star}\|+\sqrt{|\mathcal{O}(\mathscr{P}_l)\backslash(\mathcal{N}_i\cup\{i\})|}\|K_i\|W_{T_1}.
	\end{align}
	Then, defining $\rho_{li}=\sqrt{2}\|K_i\|$, and $\kappa_{li}=M_{\bar{u}_i}+M_{u_i}+\sqrt{|\mathcal{O}(\mathscr{P}_l)\backslash(\mathcal{N}_i\cup\{i\})|}\times\|K_i\|W_{T_1}$ yield the conclusion.    
\end{IEEEproof}

This section has designed a distributed-observer-based distributed control law and analyzed the influence of model mismatch and fusion estimation in distributed observer on the control input. Since the local observer no longer reconstructs all the states of the interconnected system, there is a mismatch between $\bar{u}_i$ and $\hat{u}_{ii}$, as well as $\bar{u}_i$ and $\hat{u}_{li}$. Therefore, the error forms $\bar{u}_i-\hat{u}_{ii}$ and $\bar{u}_i-\hat{u}_{li}$ in this paper are difficult to express explicitly, which is different from the situation of the traditional distributed observer. Lemma \ref{lem1} and Lemma \ref{lem2} in this subsection effectively solve this problem by introducing finite time constraints. It is worth emphasizing that the finite time condition only plays an auxiliary role in this paper, and it will be abandoned in subsection \ref{sec4.3}.


\subsection{Performance of cover-based distributed observer}\label{sec4.2}
This subsection mainly focuses on the stability of error dynamics of cover-based distributed observers. Before that, we need to provide an important lemma. Denote $e_{\star i}^{(p)}=col\{e_{ji}^{(p)},~j\in\mathcal{O}_p\}$ and then we can state the follows.
\begin{lem}\label{estar}
	Consider $e_{i\star}$ defined in Lemma \ref{lem1} and $e_{\star i}^{(p)}$. We have $\sum_{i=1}^N\sum_{\mathcal{O}_p\in\mathcal{P}_i}\|e_{\star i}^{(p)}\|^2=\sum_{i=1}^N\|e_{i\star}\|^2=\|e\|^2$, and $\sum_{i=1}^N\sum_{\mathcal{O}_p\in\mathcal{P}_i}\|e_{\star i}^{(p)}\|\leq\sqrt{2}\|e\|$, where $e=col\{e_{i\star},~i=1,\dots,N\}$.
\end{lem}
\begin{IEEEproof}
	We know $\|e_{i\star}\|^2=\sum_{\mathcal{O}_p\in\mathscr{P}_i}\sum_{j\in\mathcal{O}_p}\|e_{ij}^{(p)}\|^2$ and $\|e_{\star i}^{(p)}\|^2=\sum_{j\in\mathcal{O}_p}\|e_{ji}^{(p)}\|^2$ based on the definition of $e_{i\star}$ and $e_{\star i}^{(p)}$. Hence,
	\begin{align*}
		&\sum_{i=1}^N\sum_{\mathcal{O}_p\in\mathcal{P}_i}\left\|e_{\star i}^{(p)}\right\|^2=\sum_{i=1}^N\sum_{\mathcal{O}_p\in\mathcal{P}_i}\sum_{j\in\mathcal{O}_p}\left\|e_{ji}^{(p)}\right\|^2\\
		=&\sum_{i=1}^N\sum_{\mathcal{O}_p\in\mathcal{P}_i}\sum_{j\in\mathcal{O}_p}\left\|e_{ij}^{(p)}\right\|^2=\sum_{i=1}^N\|e_{i\star}\|^2=\|e\|^2.
	\end{align*}
	Furthermore, by the inequality of arithmetic and geometric means, we have
	\begin{align*}
		&\left(\sum_{i=1}^N\sum_{\mathcal{O}_p\in\mathcal{P}_i}\left\|e_{\star i}^{(p)}\right\|\right)^2\\
		\leq& 2\sum_{i=1}^N\sum_{\mathcal{O}_p\in\mathcal{P}_i}\left\|e_{\star i}^{(p)}\right\|^2
		=2\sum_{i=1}^N\|e_{i\star}\|^2=2\|e\|^2.
	\end{align*}
	Therefore, $\sum_{i=1}^N\sum_{\mathcal{O}_p\in\mathcal{P}_i}\|e_{\star i}^{(p)}\|\leq\sqrt{2}\|e\|$. 
\end{IEEEproof}

Now, the main theorem of this subsection can be given. Note that this is a preliminary conclusion on the stability of error dynamics of cover-based distributed observers. The complete conclusion will be presented in Theorem \ref{thm3} in the next subsection.

\begin{thm}\label{thm1}
	Consider system (\ref{sys11}), distributed observer (\ref{observer1})--(\ref{observer2}), and networks subject to Assumption \ref{assume1} and \ref{assume2}, and also consider the constant $T_1>0$ used in Lemma \ref{lem1} and \ref{lem2}. Assume that states of closed-loop system (\ref{sys21}) and (\ref{u3}) are bounded when $t<T_1$. Then, for arbitrary $T_1'<T_1$, there is a $\theta$ such that the error dynamics of (\ref{observer1})--(\ref{observer2}) can converge to an any small invariant set (around the origin) during $t<T_1'<T_1$ if
	
	1)~$H_{i}$ is chosen as $\theta^{n-2}\bar{H}_i$ such that $\bar{A}_{ii}-\bar{H}_{i}\bar{C}_{i}$ is a Hurwitz matrix (see more detail of $\bar{H}_i$ in Remark \ref{H}) for all $i=1,\ldots,N$ where $\bar{A}_{ii}=\frac{1}{\theta^{n-1}}\Gamma_\theta A_{ii}\Gamma_\theta^{-1}$ and $\bar{C}_i=C_i\Gamma_\theta^{-1}$; 
	
	2)~$P_{i}$ is a symmetric positive definite matrix solved by
	\begin{align}\label{1-C2}
		sym\{P_{i}(\bar{A}_{i}-\bar{H}_{i}\bar{C}_{i})\}=-2\gamma I_n,
	\end{align}
	
	3)~For all $\theta\geq 1$, coupling gain $\gamma$ satisfies
	\begin{align}\label{1-C3}
		\gamma>\frac{1}{2\theta^{n-1}\underline{\lambda}_{\mathcal{O}_p}}\left(\lambda_A+2\lambda_P\bar{\lambda}\left(\|A\|+\rho\|B\|\right)\right),
	\end{align}
	where  $\underline{\lambda}_{\mathcal{O}_p}=\min_{i=1,\ldots,N}\{\lambda_{\mathcal{O}_p},~\mathcal{O}_p\in\mathscr{P}_i\}$ with $\lambda_{\mathcal{O}_p}=\underline{\sigma}(\mathcal{L}_{\mathcal{O}_p}+\mathcal{S}_{\mathcal{O}_p})$, $\mathcal{L}_{\mathcal{O}_p}$ being the Laplacian matrix of the subgraph among $\mathcal{O}_p$ as well as $\mathcal{S}_{\mathcal{O}_p}=diag\{1,0,\ldots,0\}\in\mathbb{R}^{|\mathcal{O}|_p\times |\mathcal{O}|_p}$; $\lambda_{A_i}=\max_{i}\{\bar{\sigma}(A_{ii}+A_{ii}^T)\}$, $\lambda_P=\max_i\{\bar{\sigma}(P_i)\}$, $\bar{\lambda}=\max_i\{\lambda_i\}$ with $\lambda_i=|\mathscr{P}_i|\max\{|\mathcal{O}_p|,~p\in\mathscr{P}_i\}$, and $\rho=\max_i\{\rho_{i}\}$ with $\rho_{i}=\max\{\rho_{li},~l\in\mathcal{O}(\mathscr{P}_i)\}$.
\end{thm}
\begin{IEEEproof}
	Set $\eta_{ii}^{(p)}=\Gamma_\theta e_{ii}^{(p)}$ and $\eta_{ji}^{(p)}=e_{ji}^{(p)}$, then, based on equation (\ref{e1}), the dynamics of $\eta_{ii}^{(p)}$ gives rise to
	\begin{align*}
		\dot{\eta}_{ii}^{(p)}=&\theta^{n-1}\left(\bar{A}_{ii}-\bar{H}_i\bar{C}_i\right)\eta_{ii}^{(p)}+\Gamma_\theta\sum_{j\in\mathcal{N}_i}A_{ij}\bar{\eta}_{ij}\notag\\
		&+\Gamma_\epsilon^{-1} B_i(\hat{u}_{ii}-\bar{u}_i)-\gamma\theta^{n-1} P_i^{-1}\sum_{j\in\mathcal{O}_p}\alpha_{ij}\left(\eta_{ji}^{(p)}-\eta_{ii}^{(p)}\right),
	\end{align*}
	where $\bar{\eta}_{ij}=\Gamma_\theta\bar{e}_{ij}$. Denote $\eta_{\star i}^{(p)}=col\{\eta_{ji}^{(p)},~j\in\mathcal{O}_p\}$ and $\eta_{i\star}=col\{\eta_{ij},~j\in\mathcal{O}_p,~\mathcal{O}_p\in\mathscr{P}_i\}$, then
	\begin{align*}
		\dot{\eta}_{\star i}^{(p)}=&A_{\star i}^{(p)}\eta_{\star i}^{(p)}+\Gamma_{\star i}^{(p)}\Delta_{\star i}^{(p)}+\Gamma_{\star i}^{(p)}U_{\star i}^{(p)}\notag\\
		&-\gamma\theta^{n-1} \left(P_{\star i}^{(p)}\right)^{-1}\left(\mathcal{L}_{\mathcal{O}_p}\otimes I_n\right)\eta_{\star i}^{(p)}, 
	\end{align*}
	where 
	\begin{align*}
		&A_{\star i}^{(p)}=diag\left\{\theta^{n-1}(\bar{A}_{ii}-\bar{H}_i\bar{C}_i),~\underbrace{A_{ii},\ldots, A_{ii}}_{|\mathcal{O}_p|-1}\right\},\\
		&\Gamma_{\star i}^{(p)}=diag\left\{\Gamma_\theta,~\underbrace{I_n,\ldots, I_n}_{|\mathcal{O}_p|-1}\right\},\\
		&\Delta_{\star i}^{(p)}=col\left\{\Delta_{ii}^{(p)},~\Delta_{li}^{(p)},~l\in\mathcal{O}_p\right\},\\
		&\Delta_{ii}^{(p)}=\sum_{j\in\mathcal{N}_i}A_{ij}\bar{\eta}_{ij},\\
		&\Delta_{li}^{(p)}=\sum_{j\in\mathcal{N}_i\cap\mathcal{O}(\mathcal{P}_l)}A_{ij}\bar{\eta}_{lj}-\sum_{j\in\mathcal{N}_i\backslash\mathcal{O}(\mathcal{P}_l)}A_{ij}x_{j},\\
		&U_{\star i}^{(p)}=col\left\{U_{ii},~U_{li},~l\in\mathcal{O}_p\right\},\\
		&U_{ii}=B_i(\hat{u}_{ii}-\bar{u}_i),~U_{li}=B_i(\hat{u}_{li}-\bar{u}_i),\\
		&P_{\star i}^{(p)}=diag\left\{P_i,~\underbrace{I_n,\ldots, I_n}_{|\mathcal{O}_p|-1}\right\}.
	\end{align*}
	
	To move on, we have
	\begin{align}\label{delta}
		\left\|\Delta_{\star i}^{(p)}\right\|\leq& \sum_{l\in\mathcal{O}_p}\left\|\Delta_{li}^{(p)}\right\|\notag\\
		\leq&\sum_{l\in\mathcal{O}_p}\|A_i\|\left\|\eta_{l\star}\right\|+\sum_{l\in\mathcal{O}_p,~l\neq i}\|A_i\|\left\|x\right\|\notag\\
		\leq&\sum_{l\in\mathcal{O}_p}\|A_i\|\left(\left\|\eta_{l\star}\right\|+W_{T_1}\right),
	\end{align}
	where $A_i=[A_{i1},\ldots,A_{iN}]$, and $W_{T_1}$ is defined in Lemma \ref{lem2}. Besides, 
	\begin{align}\label{U}
		\left\|U_{\star i}^{(p)}\right\|\leq&\sum_{l\in\mathcal{O}_p}\|U_{li}\|\notag\\
		\leq&\|B_i\|\sum_{l\in\mathcal{O}_p}\left(\rho_{li}\|e_{l\star}\|+\kappa_{li}\right)\notag\\
		\leq&\rho_{i}\|B_i\|\sum_{l\in\mathcal{O}_p}\|e_{l\star}\|+\|B_i\||\mathcal{O}_p|\kappa_i,
	\end{align}
	where $\kappa_i=\max_{l\in\mathcal{O}_p}\{\kappa_{li}\}$. Subsequently, the Lyapunov candidate can be chosen as $V=\sum_{i=1}^N V_i$, where
	\begin{align*}
		V_i=\sum_{\mathcal{O}_p\in\mathscr{P}_i}\left(\eta_{\star i}^{(p)}\right)^TP_{\star i}^{(p)}\eta_{\star i}^{(p)}.
	\end{align*}
	Then, based on (\ref{delta}) and (\ref{U}), the derivative of $V_i$ along with $\eta_{\star i}^{(p)}$ gives rise to
	\begin{align}\label{ly1}
		\dot{V}_i=&\sum_{\mathcal{O}_p\in\mathcal{P}_i}\left(\eta_{\star i}^{(p)}\right)^Tsym\left\{P_{\star i}^{(p)}A_{\star i}^{(p)}\right\}\eta_{\star i}^{(p)}\notag\\
		&+\sum_{\mathcal{O}_p\in\mathcal{P}_i}\left(2P_{\star i}^{(p)}\Delta_{\star i}^{(p)}\eta_{\star i}^{(p)}+2P_{\star i}^{(p)}U_{\star i}^{(p)}\eta_{\star i}^{(p)}\right)\notag\\
		&-2\sum_{\mathcal{O}_p\in\mathcal{P}_i}\gamma\theta^{n-1}\left(\eta_{\star i}^{(p)}\right)^T\left(\mathcal{L}_{\mathcal{O}_p}\otimes I_n\right)\eta_{\star i}^{(p)}\notag\\
		\leq&\sum_{\mathcal{O}_p\in\mathcal{P}_i}\theta^{n-1}\left(\eta_{ii}^{(p)}\right)^Tsym\left\{P_i\left(\bar{A}_{ii}-\bar{H}_i\bar{C}_{i}\right)\right\}\eta_{ii}^{(p)}\notag\\
		&-2\sum_{\mathcal{O}_p\in\mathcal{P}_i}\gamma\theta^{n-1}\left(\eta_{\star i}^{(p)}\right)^T\left(\mathcal{L}_{\mathcal{O}_p}\otimes I_n\right)\eta_{\star i}^{(p)}\notag\\
		&+\sum_{\mathcal{O}_p\in\mathcal{P}_i}\sum_{j\in\mathcal{O}_p,~j\neq i}\left(\eta_{ji}^{(p)}\right)^Tsym\{A_{ii}\}\eta_{ji}^{(p)}\notag\\
		&+2\bar{\sigma}(P_i)\sum_{\mathcal{O}_p\in\mathcal{P}_i}\left(\sum_{l\in\mathcal{O}_p}\|A_i\|\left(\left\|\eta_{l\star}\right\|+W_{T_1}\right)\right.\notag\\
		&+\left.\rho_{i}\|B_i\|\sum_{l\in\mathcal{O}_p}\|\eta_{l\star}\|+\|B_i\||\mathcal{O}_p|\kappa_i\right)\left\|\eta_{\star i}^{(p)}\right\|.
	\end{align}
	According to conditions 1), 2), and Lemma \ref{spanning}, we have
	\begin{align}\label{ly2}
		&\left(\eta_{ii}^{(p)}\right)^Tsym\left\{P_i\Lambda_i\right\}\eta_{ii}^{(p)}-2\gamma\left(\eta_{\star i}^{(p)}\right)^T\left(\mathcal{L}_{\mathcal{O}_p}\otimes I_n\right)\eta_{\star i}^{(p)}\notag\\
		=&-2\gamma\left(\eta_{ii}^{(p)}\right)^T\eta_{ii}^{(p)}-2\gamma\left(\eta_{\star i}^{(p)}\right)^T\left(\mathcal{L}_{\mathcal{O}_p}\otimes I_n\right)\eta_{\star i}^{(p)}\notag\\
		=&-2\gamma\left(\eta_{\star i}^{(p)}\right)^T\left(\left(\mathcal{L}_{\mathcal{O}_p}+\mathcal{S}_{\mathcal{O}_p}\right)\otimes I_n\right)\eta_{\star i}^{(p)}\notag\\
		=&-2\gamma\lambda_{\mathcal{O}_p}\left\|\eta_{\star i}^{(p)}\right\|^2,
	\end{align}
	where $\Lambda_i=\bar{A}_{ii}-\bar{H}_i\bar{C}_{i}$. Then, substituting (\ref{ly2}) into (\ref{ly1}) yields
	\begin{align*}
		\dot{V}_i\leq& -2\gamma\theta^{n-1}\sum_{\mathcal{O}_p\in\mathcal{P}_i}\lambda_{\mathcal{O}_p}\left\|\eta_{\star i}^{(p)}\right\|^2+\sum_{\mathcal{O}_p\in\mathcal{P}_i}\lambda_{A_i}\left\|\eta_{\star i}^{(p)}\right\|^2\notag\\
		&+2\bar{\sigma}(P_i)\sum_{\mathcal{O}_p\in\mathcal{P}_i}\sum_{l\in\mathcal{O}_p}\left(\|A_i\|+\rho_{i}\|B_i\|\right)\left\|\eta_{l\star}\right\|\left\|\eta_{\star i}^{(p)}\right\|\notag\\
		&+2\bar{\sigma}(P_i)\sum_{\mathcal{O}_p\in\mathcal{P}_i}\left(\|A_i\|W_{T_1}+\|B_i\|\kappa_i\right)|\mathcal{O}_p|\left\|\eta_{\star i}^{(p)}\right\|\notag\\
		\leq&-2\gamma\theta^{n-1}\sum_{\mathcal{O}_p\in\mathcal{P}_i}\lambda_{\mathcal{O}_p}\left\|\eta_{\star i}^{(p)}\right\|^2+\sum_{\mathcal{O}_p\in\mathcal{P}_i}\lambda_{A_i}\left\|\eta_{\star i}^{(p)}\right\|^2\notag\\
		&+\bar{\sigma}(P_i)\left(\|A_i\|+\rho_{i}\|B_i\|\right)\sum_{\mathcal{O}_p\in\mathcal{P}_i}\sum_{l\in\mathcal{O}_p}\left\|\eta_{l\star}\right\|^2\notag\\
		&+\bar{\sigma}(P_i)\left(\|A_i\|+\rho_{i}\|B_i\|\right)\sum_{\mathcal{O}_p\in\mathcal{P}_i}\sum_{l\in\mathcal{O}_p}\left\|\eta_{\star i}^{(p)}\right\|^2\notag\\
		&+2\bar{\sigma}(P_i)\left(\|A_i\|W_{T_1}+\|B_i\|\kappa_i\right)\sum_{\mathcal{O}_p\in\mathcal{P}_i}|\mathcal{O}_p|\left\|\eta_{\star i}^{(p)}\right\|.
	\end{align*}
	
	Note that $\sum_{i=1}^N\sum_{\mathcal{O}_p\in\mathcal{P}_i}\sum_{l\in\mathcal{O}_p}\|\eta_{l\star}\|^2$ is equivalent to repeatedly calculating $\sum_{i=1}^N\|\eta_{i\star}\|^2$ for all $i=1,\ldots,N$ up to $\bar{\lambda}$ times and so does $\sum_{i=1}^N\sum_{\mathcal{O}_p\in\mathcal{P}_i}\sum_{l\in\mathcal{O}_p}\|\eta_{\star i}^{(p)}\|^2$, where $\bar{\lambda}=\max_i\{\lambda_i\}$ with $\lambda_i=|\mathscr{P}_i|\max\{|\mathcal{O}_p|,~p\in\mathscr{P}_i\}$. It leads thereby to
	\begin{align*}
		\dot{V}\leq&\left(-2\gamma\theta^{n-1}\underline{\lambda}_{\mathcal{O}_p}+\lambda_{A_i}\right)\sum_{i=1}^N\sum_{\mathcal{O}_p\in\mathcal{P}_i}\left\|\eta_{\star i}^{(p)}\right\|^2\notag\\
		&+\lambda_P\left(\|A\|+\rho\|B\|\right)\sum_{i=1}^N\sum_{\mathcal{O}_p\in\mathcal{P}_i}\sum_{l\in\mathcal{O}_p}\left\|\eta_{l\star}\right\|^2\notag\\
		&+\lambda_P\left(\|A\|+\rho\|B\|\right)\sum_{i=1}^N\sum_{\mathcal{O}_p\in\mathcal{P}_i}\sum_{l\in\mathcal{O}_p}\left\|\eta_{\star i}^{(p)}\right\|^2\notag\\
		&+2\lambda_P\left(\|A\|W_{T_1}+\|B\|\kappa\right)\sum_{i=1}^N\sum_{\mathcal{O}_p\in\mathcal{P}_i}|\mathcal{O}_p|\left\|\eta_{\star i}^{(p)}\right\|\notag\\
		\leq&\left(-2\gamma\theta^{n-1}\underline{\lambda}_{\mathcal{O}_p}+\lambda_{A_i}\right)\sum_{i=1}^N\sum_{\mathcal{O}_p\in\mathcal{P}_i}\left\|\eta_{\star i}^{(p)}\right\|^2\notag\\
		&+\lambda_P\bar{\lambda}\left(\|A\|+\rho\|B\|\right)\sum_{i=1}^N\|\eta_{i\star}\|^2\notag\\
		&+\lambda_P\bar{\lambda}\left(\|A\|+\rho\|B\|\right)\sum_{i=1}^N\sum_{\mathcal{O}_p\in\mathcal{P}_i}\left\|\eta_{\star i}^{(p)}\right\|^2\notag\\
		&+2\lambda_P\bar{\lambda}\left(\|A\|W_{T_1}+\|B\|\kappa\right)\sum_{i=1}^N\sum_{\mathcal{O}_p\in\mathcal{P}_i}\left\|\eta_{\star i}^{(p)}\right\|,
	\end{align*}
	where $\kappa=\max_{i=1,\ldots,N}\{\kappa_{i}\}$. Note that Lemma \ref{estar} indicates $$\sum_{i=1}^N\sum_{\mathcal{O}_p\in\mathcal{P}_i}\|\eta_{\star i}^{(p)}\|^2=\sum_{i=1}^N\|\eta_{i\star}\|^2,$$ and $$\sum_{i=1}^N\sum_{\mathcal{O}_p\in\mathcal{P}_i}\|\eta_{\star i}^{(p)}\|\leq\sqrt{2}\|\eta\|,$$ where $\eta=col\{\eta_{i\star},~i=1,\dots,N\}$. Then, by denoting $c(\theta)=2\gamma\theta^{n-1}\underline{\lambda}_{\mathcal{O}_p}-\lambda_{A_i}-2\lambda_P\bar{\lambda}\left(\|A\|+\rho\|B\|\right)$, and $c_K=2\lambda_P\bar{\lambda}\left(\|A\|W_{T_1}+\|B\|\kappa\right)$, we can further obtain
	\begin{align}\label{ly4}
		\dot{V}\leq&-c(\theta)\|\eta\|^2+c_K\|\eta\|.
	\end{align}
	Condition 3) indicates that $c(\theta)>0$, and $c(\theta)$ is obviously monotonically increasing and radially unbounded with respect to $\theta^{n-1}$. 
	
	Equation (\ref{ly4}) infers that $\Omega_{\eta}=\{\|\eta\|:~\|\eta\|\leq c_K/c(\theta)\}$ is an invariant set. So, $\dot{V}<0$ when $\|\eta\|$ is out of $\Omega_{\eta}$. Then, we obtain by the conclusion of \cite{7930554} that
	\begin{align}
		\|\eta(t)\|\leq \max\left\{be^{-ac(\theta)t}\|\eta(0)\|,~\frac{c_K}{c(\theta)}\right\}
	\end{align}
	for all $t>0$ and some positive constants $a,b$. It means
	\begin{align}
		\|e(t)\|\leq \max\left\{b\theta^{n-1}e^{-ac(\theta)t}\|e(0)\|,~\frac{c_K}{c(\theta)}\right\},
	\end{align}
	which shows that $\|e\|$ can converge to an any small invariant set $\Omega_{e}=\{\|e\|:~\|e\|\leq c_K/c(\theta)\}$ in any short time by designing proper $\theta$. In other words, one can choose a proper $\theta$ so that $\|e\|$ converges to $\Omega_{e}$ for any $T_1'<T_1$.  
\end{IEEEproof}

This section has proved that error dynamics of the cover-based distributed observer can converge to an any small invariant set before the given time $T_1$. Due to the inability to analyze the performance of distributed observers and closed-loop systems separately, Theorem \ref{thm1} relies on assumptions about $T_1$. In Theorem \ref{thm3} of the next subsection, we will provide a joint analysis of the observer and controller, where the assumption about $T_1$ will be removed. Next, we will further elaborate on some details and supplementary explanations of Theorem \ref{thm1} in several remarks.

\begin{rem}
	The design methods for the parameters of the cover-based distributed observer are detailed in Theorem \ref{thm1}. According to Theorem \ref{thm1}, all parameters except for $\gamma$ only rely on the information of each individual agent. Though designing $\gamma$ involves the information of system matrix, input matrix, and network topology, $\gamma$ is actually just a sufficiently large constant. Therefore, in practical usage, it is often avoided to use global information by designing $\gamma$ as an adaptive parameter \cite{2019Completely,Xu2025TAC}.
\end{rem}


\begin{rem}
	Fusion estimation and model mismatch in cover-based distributed observer design make the stability analysis of its error dynamics (\ref{e1})--(\ref{e2}) a real challenge because they not only bring more complex error dynamics but also derive two distinct compact error forms ($e_{\star i}^{(p)}$ and $e_{i\star}$). In this subsection, Lemma \ref{estar} demonstrates the strict mathematical relationship between these two compact error forms. Then, Theorem \ref{thm1} develops the two-layer Lyapunov analysis method, which ingeniously transforms the above mathematical relationship into the results that can be used in the traditional Lyapunov stability analysis. The proof of Lemma \ref{estar} and the design of the two-layer Lyapunov function ($V=\sum_{i=1}^NV_i$) are the keys to successfully proving the error dynamic stability of the cover-based distributed observer.
\end{rem}

\subsection{Closed-loop performance under the distributed control law}\label{sec4.3}

This subsection first proves the performance of the closed-loop system under the assumption of stable observer error (Theorem \ref{thm2}). Then, we presents the joint analysis results of the observer and controller in Theorem \ref{thm3}, in which the stability analysis of the error dynamics of the distributed observer and the closed-loop system dynamics is completed (without relying on additional assumptions). 

At the beginning, one obtains that the closed-loop system with centralized control law (\ref{u1}) is
\begin{align}\label{xc}
	\dot{x}=Ax+BKx=(A+BK)x.
\end{align}
Furthermore, the closed-loop system under distributed control law (\ref{u2}) takes the form of
\begin{align}\label{xr-1}
	\dot{x}_i=&A_{ii}x_i+\sum_{j\in\mathcal{N}_i}A_{ij}x_j+B_i\sum_{j\in\mathcal{N}_i\cup\{i\}}K_{ij}\bar{x}_{ij}\notag\\
	=&A_{ii}x_i+\sum_{j\in\mathcal{N}_i}A_{ij}x_j+B_i\sum_{j\in\mathcal{N}_i\cup\{i\}}K_{ij}x_{j}\notag\\
	&+B_i\sum_{j\in\mathcal{N}_i\cup\{i\}}K_{ij}\bar{x}_{ij}-B_i\sum_{j\in\mathcal{N}_i\cup\{i\}}K_{ij}x_{j}.
\end{align}
Let $U=col\{\hat{u}_{ii}-u_i,~i=1,\ldots,N\}$, then the compact form of (\ref{xr-1}) is expressed as
\begin{align}\label{xr}
	\dot{x}=(A+BK)x+BU.
\end{align}
Now, we states the follows to show the performance of (\ref{xr}).

\begin{thm}\label{thm2}
	Consider the closed-loop system (\ref{xr}) as well as its communication network and physical network subject to Assumption \ref{assume1} and \ref{assume2}. If the error $\|e\|$ of the cover-based distributed observer (\ref{observer1})--(\ref{observer2}) stays in $\Omega_e$ with $t\geq 0$, then $\|x\|$ can converge to an invariant 
	\begin{align*}
		\Omega_x=\{\|x\|:~\|x\|\leq 4c_K\|QB\|\|K\|/c(\theta)c_2\}.
	\end{align*}
	where $c(\theta)=\gamma\theta\underline{\lambda}_{\mathcal{O}_p}-\lambda_A/\theta^{n-1}-2\lambda_P\bar{\lambda}\left(\|A\|+\rho\|B\|\right)$---given in the proof of Theorem \ref{thm1}---is a monotonically increasing function with respect to $\theta$ and radially unbounded. Furthermore, $\Omega_x$ can be  arbitrarily small with observer parameter $\theta$ tending to infinity.
\end{thm}
\begin{IEEEproof}
	Since $A+BK$ is Hurwitz, there is a symmetric positive definite matrix $Q$ so that $sym\{Q(A+BK)\}=-c_2I_{nN}$, where $c_2>0$ is a given constant. Then, the Lyapunov candidate can be chosen as $W(t)=x^TQx$. Its derivative along with (\ref{xr}) gives rise to
	\begin{align}
		\dot{W}=x^Tsym\{Q(A+BK)\}x+2x^TQBU.
	\end{align}
	Based on (\ref{uii-ui}), we know $\|\hat{u}_{ii}-u_i\|\leq\sqrt{2}\|K_i\|\|e_{i\star}\|$. Hence,
	\begin{align}
		\dot{W}\leq& -c_2\|x\|^2+2\|x\|\|QB\|\sum_{i=1}^N\sqrt{2}\|K_i\|\|e_{i\star}\|\notag\\
		\leq&-c_2\|x\|^2+2\sqrt{2}\|x\|\|QB\|\|K\|\sum_{i=1}^N\|e_{i\star}\|\notag\\
		\leq&-c_2\|x\|^2+4\|QB\|\|K\|\|x\|\|e\|\notag\\
		\leq&-c_2\|x\|^2+\frac{4c_K}{c(\theta)}\|QB\|\|K\|\|x\|.
	\end{align}
	It means that $\Omega_x=\{\|x\|:~\|x\|\leq 4c_K\|QB\|\|K\|/c(\theta)c_2\}$ is an invariant set. Furthermore, since $c(\theta)$ is a monotonically increasing function with respect to $\theta$ and radially unbounded, we have
	\begin{align}
		\lim_{\theta\to\infty}\frac{4c_K\|QB\|\|K\|}{c(\theta)c_2}=0.
	\end{align}
	Therefore, $\Omega_x$ is an arbitrary small invariant. 
\end{IEEEproof}

In what follows, we will focus on the closed-loop system with the distributed control (\ref{u4}) containing saturation mechanism:
\begin{align}
	\dot{x}_i=A_{ii}x_i+\sum_{j\in\mathcal{N}_i}A_{ij}x_j+B_i\sum_{j\in\mathcal{N}_i\cup\{i\}}K_{ij}\mathbbm{x}_{ij}.
\end{align}
This is the actual control law adopted by the closed-loop system in this paper, and also the source of output information $y_i$ in the cover-based distributed observer (\ref{observer1})--(\ref{observer2}). Let $\bar{U}=col\{\bar{u}_i-u_i,~i=1,\ldots,N\}$ and obtain the compact form
\begin{align}\label{xr-s}
	\dot{x}=(A+BK)x+B\bar{U}.
\end{align}
The following theorem will show the stability of (\ref{xr-s}).
\begin{thm}\label{thm3}
	The distributed-observer-based distributed control law for large-scale system is given by (\ref{observer1})--(\ref{observer2}) and (\ref{u2})--(\ref{u4}). If we choose proper gains $K_{ij}$ in (\ref{u2}) and observer parameters by Theorem \ref{thm1}, then the states of the closed-loop system (\ref{xr-s}) can converge to an invariant set
	\begin{align}
		\Omega_x=\{\|x\|:~\|x\|\leq 4c_K\|QB\|\|K\|/c(\theta)c_2\}.
	\end{align} 
	This set can be arbitrarily small when observer parameter $\theta$ tending to infinity.
\end{thm}
\begin{IEEEproof}
	As a linear system, there is no finite time escape problem in the system (\ref{xr-s}). Hence, there are constants $T_1>0$ and $W_{T_1}>0$ so that $\|x\|\leq W_{T_1}$ for all $0\leq t\leq T_1$. Then, in light of Theorem \ref{thm1}, we know there is a proper $\theta>0$ such that $\|e\|$ converges to $\Omega_c$ during $(0,T_1']$ where $T_1'$ is a constant satisfying $0<T_1'<T_1$.
	
	Subsequently, at time $T_1'$, we have $\|x\|\leq W_{T_1}$, which yields $\|\bar{x}_{ij}\|\leq\|x_j\|+\|\bar{e}_{ij}\|\leq W_{T_1}+c_K/c(\theta)$. By choosing $\mathcal{M}>W_{T_1}+c_K/c(\theta)$ for all $\theta\geq 1$, we know $\bar{u}_i=\hat{u}_{ii}$ at $t=T_1'$. In other words, the closed-loop system (\ref{xr-s}) degenerates into the closed-loop system (\ref{xr}).
	
	Let $\Omega_W=\{\|x\|:~\|x\|\leq W_{T_1}\}$, then $(\|e\|,\|x\|)\in\Omega_c\times\Omega_W$ at $t=T_1'$. It follows by Theorem \ref{thm2} that $\dot{W}<0$ if $\|x\|\notin\Omega_x$. Therefore, $\|x\|$ will not leave $\Omega_W$ and thus it will always be bounded by $W_{T_1}$, which infers that $\|e\|$ stays in $\Omega_e$ for $t>T_1'$ owing to Theorem \ref{thm1}. As a result, $\|x\|$ will go into $\Omega_x$ because of the conclusion of Theorem \ref{thm2}. Hence, $(\|e\|,\|x\|)$ will go into $\Omega_c\times\Omega_x$ with $t\to\infty$. 
\end{IEEEproof}
\begin{figure}[!t]
	\centering
	\includegraphics[width=8cm]{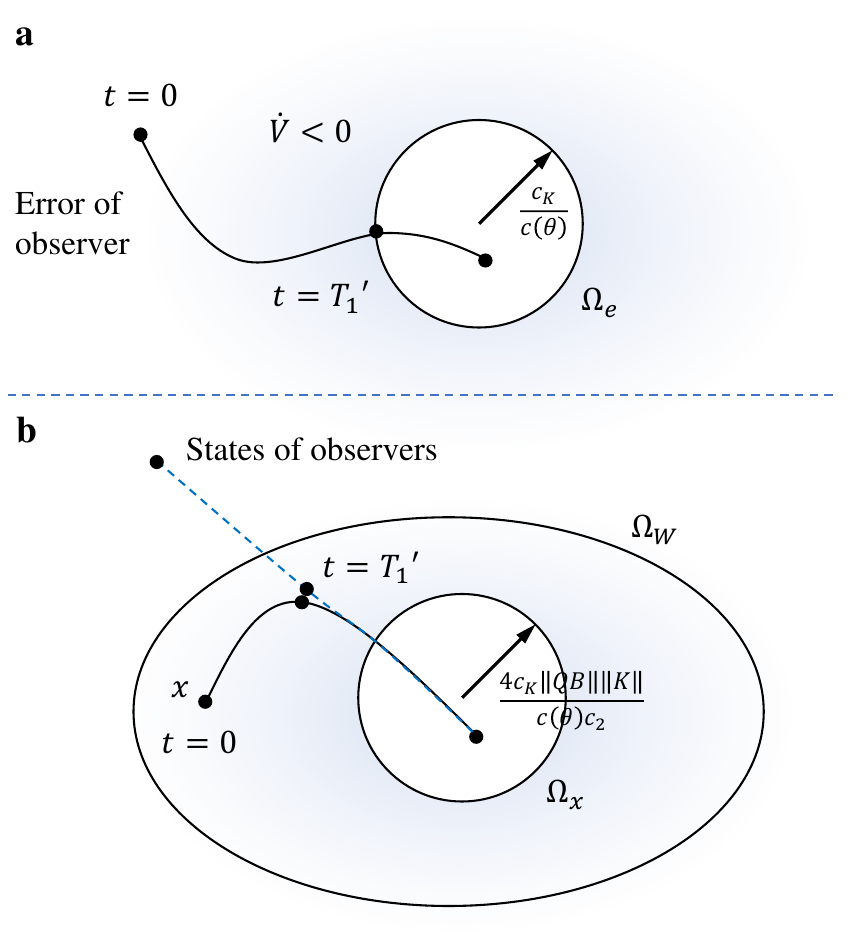}\\	
	\caption{Schematic diagram of the main proof process.}\label{proof}
\end{figure}

Theorem \ref{thm3} is of primary importance for this paper. It combines the conclusions of Theorem \ref{thm1} and Theorem \ref{thm2}, and obtains the final result: the error system and the closed-loop system of the cover-based distributed observer can converge to any small invariant sets. Furthermore, the finite time conditions involved in Theorem \ref{thm1} are eliminated by the analysis in Theorem \ref{thm3}. In fact, the final conclusion of the theorem only depends on the controllability and observability of the system as well as the connectivity of the communication network and does not depend on any other assumptions. These conclusions also indicate that, compared with the traditional distributed observer, the performance of the cover-based distributed observer in this paper has almost no loss, but the dimension of observers on each agent is greatly reduced. These results make the distributed observer more practical in large-scale system problems.

\begin{rem}
	The proof of the stability of observer error dynamics and closed-loop systems mainly involves three steps. First, we assume that the states of the closed-loop system are bounded, and then the error dynamic $e(t)$ can be proved to converge to $\Omega_e$ when $t=T_1'$ (see in Figure \ref{proof}a). Second, we show that the states of the closed-loop system can converge to $\Omega_x$ when $e_i$ is always in $\Omega_e$. Finally, as seen in Figure \ref{proof}b, we assume that $\Omega_W$ is the bound of the closed-loop system before $T_1$. By designing $\theta$, the trajectory of the observer can converge to the neighbor of the true states before $x$ escapes $\Omega_W$. After that, the distributed control law can control the dynamics of the closed-loop system to an arbitrarily small invariant set $\Omega_x$.
\end{rem}

\subsection{The ability of performance recovery}\label{sec4.4}

This subsection will show that the state trajectories of the closed-loop system (\ref{xr-s}) controlled by the distributed-observer-based distributed control law can approximate the trajectories of (\ref{xc}) arbitrarily. To this end, we denote $x_c(t)$ the solution of closed-loop system (\ref{xc}) with initial states $x_c(0)=x_0$. Moreover, the solution of (\ref{xr-s}) is defined as $x_r(t)$ with initial value $x_r(0)=x_0$. Note that the trajectories of $x_r(t)$ depends on $\theta$.

\begin{thm}
	The performance of distributed-observer-based distributed control law can approach that of the centralized control law arbitrarily, i.e.,  for any $\varepsilon>0$, there exists a constant $M>0$ and $\theta>M$, such that $\|x_r(t)-x_c(t)\|<\varepsilon$ for all $t>0$.
\end{thm}
\begin{IEEEproof}
	Since $K$ is chosen so that $A+BK$ is a Hurwitz matrix, the system (\ref{xc}) is stable. Hence, there exists $T_2>0$ such that $\|x_c(t)\|\leq\varepsilon/2$ for all $t>T_2$. Besides, according to Theorem \ref{thm3}, there is $M_1>0$ so that $4c_K\|QB\|\|K\|/c(\theta)c_2<\varepsilon/2$ for all $\theta>M_1$. Then, there is a constant $T_3>0$ such that $\|x_r(t)\|<\varepsilon/2$ for all $\theta>M_1$ and $t>T_3$. Subsequently, we have
	\begin{align}
		\|x_c(t)-x_r(t)\|\leq\|x_c(t)\|+\|x_r(t)\|<\varepsilon
	\end{align}
	for arbitrary $t>T\triangleq\max\{T_2,T_3\}$. Up to now, we have proved that $\|x_c(t)-x_r(t)\|<\varepsilon$ holds on $t\in(T,\infty)$. In what follows, we will show it also holds on $t\in(0,T]$.
	
	For this purpose, we construct a function $T(\theta)$, which satisfies $\lim_{\theta\to\infty}T(\theta)=0$ and $\|e\|<\delta'(\theta)$ for $t>T(\theta)$, where $\delta'(\theta)>0$ is a function with $\lim_{\theta\to\infty}\delta'(\theta)=0$. 
	
	We first consider the interval $(0,T(\theta)]$. Note that both $x_c(t)$ and $x_r(t)$ are bounded, as well as $\dot{x}_c(0)$ and $\dot{x}_r(0)$ are finite, thus there are two positive proportional functions $f_c(t)=k_ct$ and $f_r(t)=k_rt$ satisfying $\|x_c(t)-x_c(0)\|\leq k_ct$ and $\|x_r(t)-x_r(0)\|\leq k_rt$ when $t\in(0,T(\theta)]$, which leads thereby to
	\begin{align*}
		\|x_c(t)-x_r(t)\|&\leq\|x_c(t)-x_c(0)\|+\|x_r(t)-x_r(0)\|\notag\\
		&\leq k_ct+k_rt\leq kT(\theta)
	\end{align*}  
	for all $t\in(0,T(\theta)]$, where $k=\max\{k_c,k_r\}$. As a result, there exists a constant $M_2>0$ so that $T(\theta)<\varepsilon/k$ for arbitrary $\theta>M_2$. Therefore, we have $\|x_c(t)-x_r(t)\|<\varepsilon$ $\forall \theta>M_2$ and $t\in(0,T(\theta)]$.
	
	Finally, we consider the scenario where $t\in(T(\theta),T]$. We define $F_c(t,x)=(A+BK)x$ and $F_r(t,x,e)=(A+BK)x+B\bar{U}$, where $\bar{U}$ is an implicit function of $e$ in light of Theorem \ref{thm2} and \ref{thm3}. Since $(T(\theta),T]$ is a finite time interval, the conditions of the famous theorem named ``The Continuous Dependence Theorem of Solutions of Differential Equations on Initial Values and Parameters" \cite{hartman2002ordinary} are fulfilled. It indicates that there is a constant $\delta>0$ such that $\|x_c(t)-x_r(t)\|<\varepsilon$ if the parameter $e$ in $F_r(t,x,e)$ satisfying $\|e\|<\delta$. Bearing in mind the conclusion of Theorem \ref{thm1} and \ref{thm3}, there exists $M_3>0$ such that $\|e\|<\delta'(\theta)<\delta$ for all $\theta>M_3$. 
	
	Therefore, there exists $\theta>M=\max\{M_1,M_2,M_3\}$ so that $\|x_c(t)-x_r(t)\|<\varepsilon$, $\forall\varepsilon>0$ and $\forall t\in(0,\infty)$. 
\end{IEEEproof}

Up to now, the goals of this paper have been fully realized. After find coverage, the dimension of each local observer is far less than that of the interconnected system, but the distributed control law designed based on the cover-based distributed observer can still ensure that the controlled system can approximate the centralized control performance arbitrarily.

\section{Simulation}\label{sec5}
To illustrate the effectiveness of the developed methods, this section will be divided into three parts. Firstly, we describe the simulation system. Then, the effectiveness of the coverage solving method will be demonstrated. Finally, we show the validity of the cover-based distributed observer and the distributed-observer-based distributed control law.

\subsection{System formulation}
The frequency subsystem of the droop control system contained in the microgrid system is considered. Assume that the large-scale system contains $N$ microgrids. Let $\delta_i$ be the electrical angle of the $i$th generator, and $\omega_i$ be its angular velocity. Then, the dynamics of $(\delta_i,\omega_i)^T$ are governed by
\begin{align}
	\dot{\delta}_i=&\omega_i,\label{sim11}\\
	\dot{\omega}_i=&-\frac{1}{\tau_{P_i}}\omega_i-\frac{k_{P_i}}{\tau_{P_i}}\left(P_{1i}V_i^2+P_{2i}V_i+P_{3i}-P_d\right)\notag\\
	&-\frac{k_{P_i}}{\tau_{P_i}}\left(\sum_{j=1}^N|\beta_{ij}|V_iV_j\sin(\delta_i-\delta_j)\right)-\frac{1}{\tau_{P_i}}u_i,\label{sim12}\\
	y_i=&\delta_i,\label{sim13}
\end{align} 
where $\tau_{P_i}$ and $k_{P_i}$ represent the filter coefficient for measuring active power and frequency drop gain, respectively; $P_d$ stands for the expected active power; and $V_i$ is the voltage of the $i$th microgrid system; $\beta_{ij}$ represents the coupling relationship among microgrid $i$ and $j$. Then, by linearizing the system (\ref{sim11})--(\ref{sim13}) around the equilibrium point and denoting $x_i=(x_{i1},x_{i2})^T$ with $x_{i1}=\delta_i$ and $x_{i2}=\omega_i$, we have
\begin{align}
	\dot{x}_i=&A_{ii}x_i+\sum_{j=1}^NA_{ij}x_j+B_iu_i,\label{sim21}\\
	y_i=&C_ix_i=x_{i1},\label{sim22}
\end{align}
where 
\begin{align*}
	&A_{ii}=\begin{bmatrix}
		0&1\\-\frac{k_{P_i}}{\tau_{P_i}}\sum_{j=1}^N|\beta_{ij}|V_iV_j&-\frac{1}{\tau_{P_i}}
	\end{bmatrix},\\
	&A_{ij}=\begin{bmatrix}
		0&0\\\frac{k_{P_i}}{\tau_{P_i}}|\beta_{ij}|V_iV_j&0
	\end{bmatrix},~B_i=\begin{bmatrix}0\\1\end{bmatrix},~C_i=\begin{bmatrix}1&0\end{bmatrix}.
\end{align*}

The physical network and communication network corresponding to the system are both shown in Figure \ref{target}. $\beta_{ij}$ and $\alpha_{ij}$ are the elements of their adjacency matrices, respectively. Therefore, the large-scale system considered in this section contains $47$ subsystems, i.e., $N=47$.

In this paper, we randomly select $\tau_{P_i}$ within $[0.012,0.018]$, and randomly select $k_{P_i}$ within $[1\times 10^{-15}, 10\times 10^{-15}]$. Besides, we choose $V_i=110{\rm V}$ for all $i=1,\ldots,N$.  

\subsection{Effectiveness of coverage solving method}
It is seen from Figure \ref{target} that the physical network and communication network among $47$ agents are different. We know $S_{pc}=84.75\%$ by calculating the similarity between these two networks. 

\begin{figure}
	\centering
	\includegraphics[width=8.1cm]{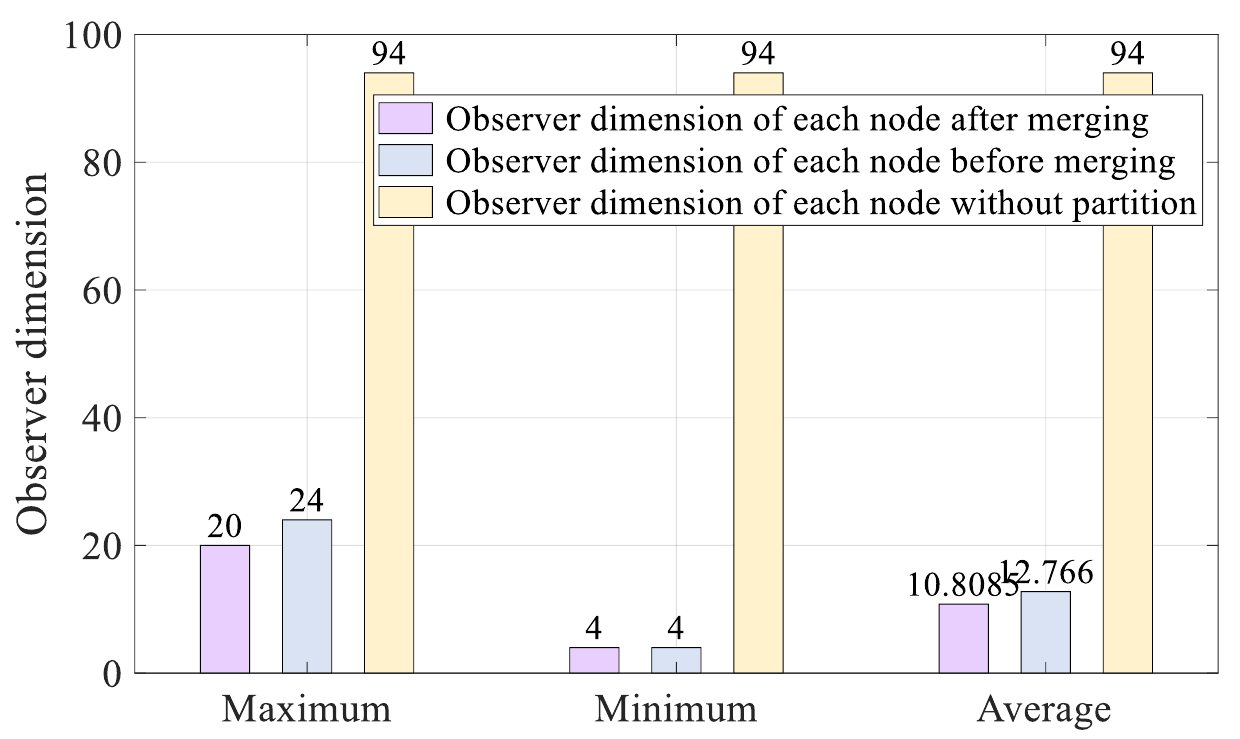}\\
	\caption{Variation of dynamic process performance with $\theta$.}\label{tongji1}
\end{figure}

Each agent needs to estimate the whole system's states if the classical distributed observer method \cite{battilotti2019distributed,2019Completely,Han2017A,Xu2021IJRNC} is used. Therefore, the dimension of observers on each agent is $94$. In contrast, it can be greatly reduced if the cover-based distributed observer developed in this paper is employed. See in Figure \ref{tongji1}; the purple bar shows that, after constructing coverage, the maximum observer dimension of all agents is only $20$ (reduced by $78.72\%$), which is much lower than that of the traditional method (yellow bar). In addition, after constructing coverage, the minimum observer dimension on each agent is only $4$ (reduced by $95.7\%$), and the average value is only $10.8083$ (reduced by $88.51\%$). Both of them are much lower than the value of the yellow bar. By the way, the merging process in Algorithm \ref{alg1} in this paper can also effectively reduce the dimension of observers on each agent. By comparing the blue with purple bars, we see that the maximum and average values of the local observer dimensions after merging cover sets (purple bar) are smaller than those before merging sets (blue bar). 

\subsection{Performance of the distributed-observer-based distributed control law}

We design the centralized control law of system (\ref{sim21}) as
\begin{align}
	u_i=\sum_{j=1}^NK_{ij}x_j+K_{ii}x_i,
\end{align}
where $K_{ij}=\begin{bmatrix}
	\frac{k_{P_i}}{\tau_{P_i}}|\beta_{ij}|V_iV_j&0
\end{bmatrix}$ and $K_{ii}$ is chosen so that $A_{ii}+B_iK_{ii}$ is a Hurwitz matrix. Subsequently, we obtain the distributed control law $\bar{u}_i$ by designing a cover-based distributed observer. 

The observer gains $H_i$, and the weighted matrix $P_i$ can be calculated by Theorem \ref{thm1} (Since there are $47$ different $H_i$ and $P_i$, we will not show their specific forms). We set $\gamma=100\theta^2$ and randomly select the initial values of the closed-loop system within $[0,1]$. In addition, the initial values of the cover-based distributed observer are all chosen as $\hat{x}_{ij}^{(p)}=2\times 1_{2|\mathcal{O}|_p}$ for all $\mathcal{O}_p\in\mathscr{P}_i$ and all $i=1,\ldots,47$.

\begin{figure}[!t]
	\centering
	\includegraphics[width=8.5cm]{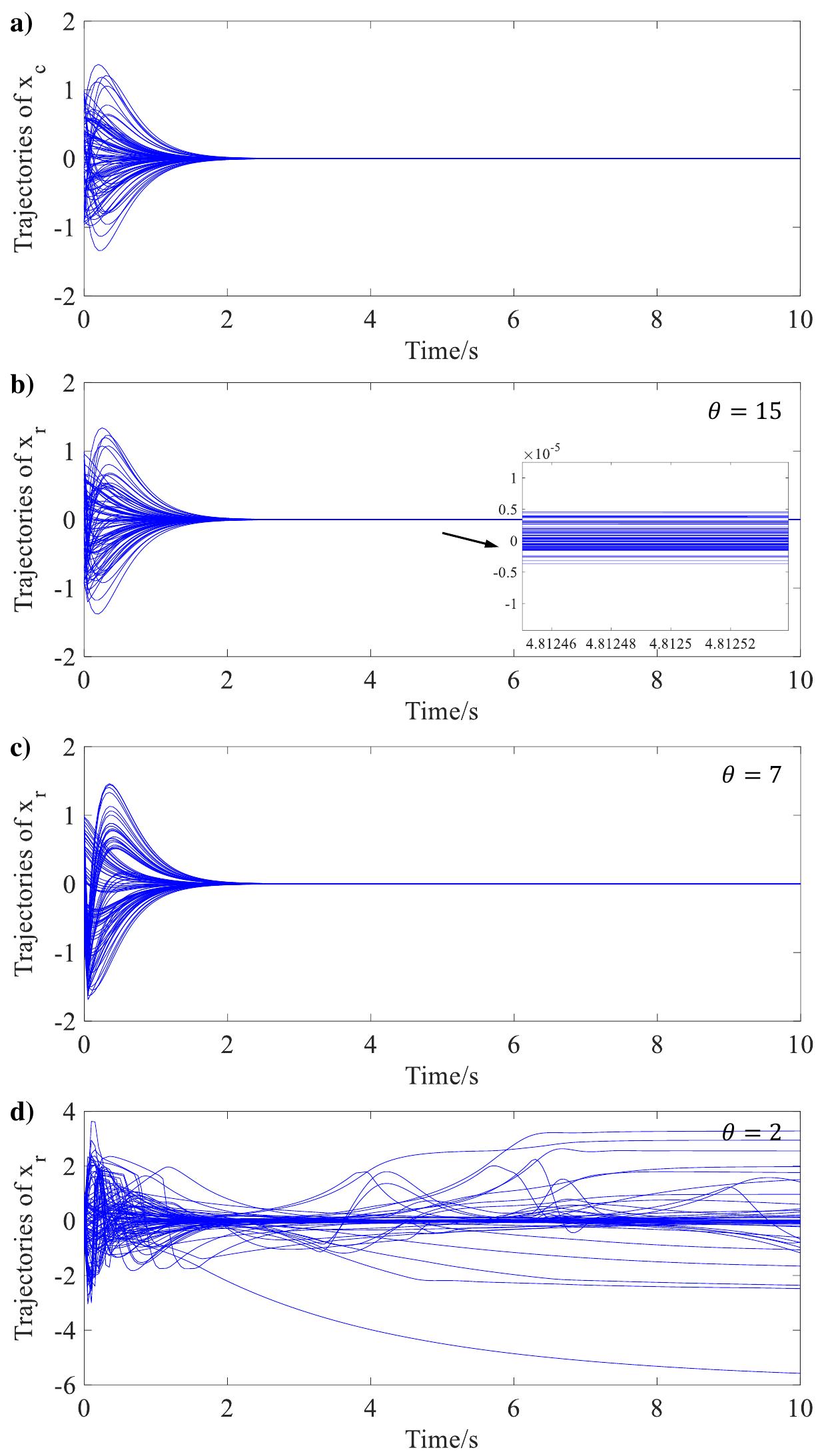}\\
	\caption{Performance of the distributed-observr-based distributed control law. a) is the trajectories of $x_c(t)$, and b), c), and d) are the trajectories of $x_r(t)$ with different $\theta$.}\label{closed-loop-tra}
\end{figure}

Then, the trajectories of the closed-loop system with different $\theta$ are shown in Figure \ref{closed-loop-tra}. Subfigure a) is the trajectories of $x_c(t)$ yielded by the centralized control law. Subfigures b), c), and d) are trajectories of $x_r(t)$, which is obtained by the distributed-observer-based distributed control law. From Figure \ref{closed-loop-tra}d) and \ref{closed-loop-tra}c), we know $x_r(t)$ cannot converge to zero when high-gain parameter $\theta=2$, while $x_r(t)$ tends to zero when $\theta=7$. However, it can be seen in Figure \ref{closed-loop-tra}c) that the dynamic performance of $x_r(t)$ with $\theta=7$ is still not as good as that of $x_c(t)$. When $\theta=15$, we find that the trajectories in Figure \ref{closed-loop-tra}b) are almost the same as that in Figure \ref{closed-loop-tra}a). It indicates that the distributed-observer-based distributed control can achieve the same performance as that of the centralized control law. 

\begin{figure}[!t]
	\centering
	\includegraphics[width=8.5cm]{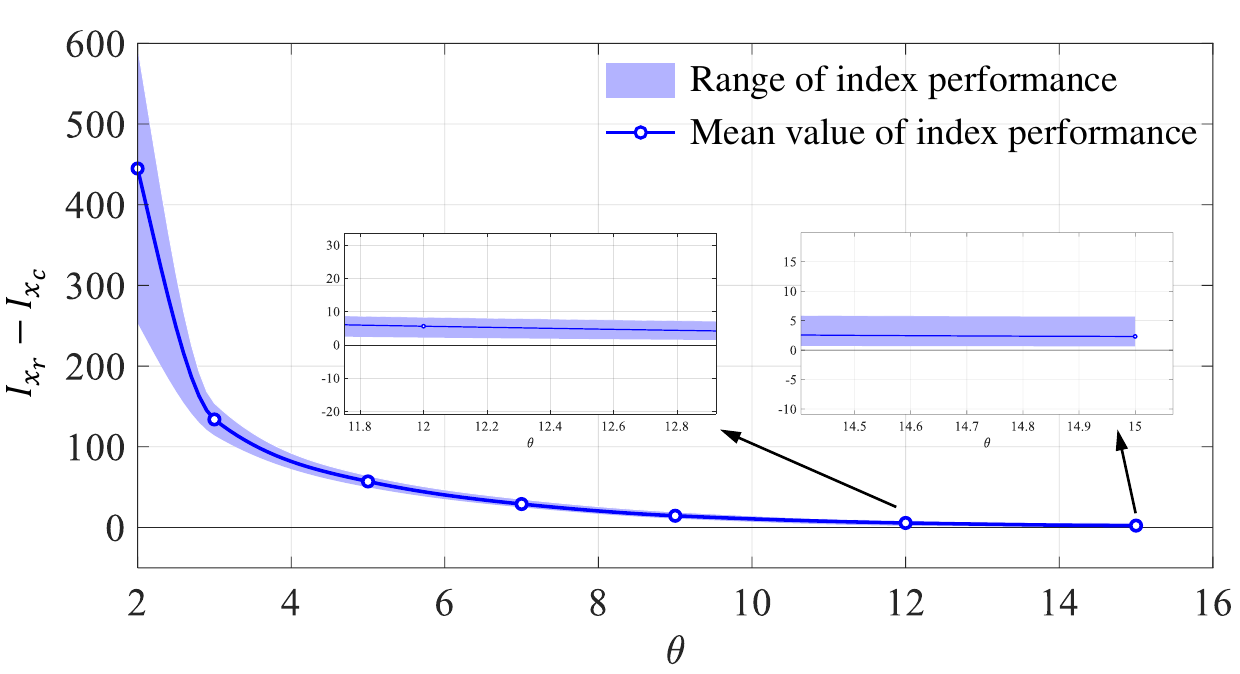}\\
	\caption{Variation of dynamic process performance with $\theta$.}\label{I-theta}
\end{figure}
In addition, we need to point out that Figure \ref{closed-loop-tra} also shows the validity of Theorem \ref{thm3}. In this figure, closed-loop system states converge to a very small invariant set (The steady-state error shown in Figure \ref{closed-loop-tra}b) is almost $5\times 10^{-6}$). In addition, the premise that the closed-loop system states can converge to any small invariant set is that the state estimation error of the cover-based distributed observer can converge to any invariant set at any fast speed. Therefore, the excellent performance of the closed-loop system in Figure \ref{closed-loop-tra}b) infers the excellent performance of the cover-based distributed observer.

To further illustrate the ability of performance recovery, we selected $\theta=2$, $\theta=3$, $\theta=5$, $\theta=7$, $\theta=9$, $\theta=12$, and $\theta=15$ to show the approximation process of distributed-observer-based distributed control performance to centralized control performance. Define the performance index of the dynamic process as
\begin{align}
	I_x=\sum_{i=1}^{47}\sum_{j=1}^2\int_0^{10}\|x_{ij}(t)\|^2dt.
\end{align}
Let $I_{x_c}$ and $I_{x_r}$ be the performance index of centralized control law and distributed control law, respectively. Then, Figure \ref{I-theta} shows how $I_{x_r}-I_{x_c}$ changes with $\theta$. Since the initial values are randomly selected, we have simulated $6$ times for each $\theta$ to eliminate the influence of the randomness of the initial values. See in Figure \ref{I-theta}, the blue dot represents the average value of $I_{x_r}-I_{x_c}$ corresponding to each $\theta$, and the blue shadow represents the floating range of $I_{x_r}-I_{x_c}$ obtained from $6$ simulations. Simulation results show that the control performance of the distributed-observer-based distributed control law is infinitely close to that of the centralized control law.

It is worth mentioning that literature \cite{Xu2022TIV} also studied the distributed-observer-based distributed control law, which can arbitrarily approximate the centralized control performance. However, it studied small-scale systems, while this paper studies large-scale systems. Furthermore, according to the developed method in \cite{Xu2022TIV}, the dimension of the local observer on each agent is equal to the dimension of the whole interconnected system. As a contrast, in this paper, the average dimension of observers on each agent is only $11.49\%$ of the interconnected system dimensions (See in Figure \ref{tongji1}).


\section{Conclusions}\label{sec6}
This paper has developed a distributed control law based on the cover-based distributed observer for large-scale interconnected linear systems. Firstly, we have designed a coverage solving algorithm, which is achieved by two steps, including initializing and merging. The algorithm can significantly reduce the dimension of local observers. Secondly, the cover-based distributed observer for large-scale systems has been designed. To analyze the stability of its error dynamics, this paper has proposed the two-layer Lyapunov analysis method and proved the dynamic transformation lemma of compact errors. Finally, we have designed the distributed control law based on the developed cover-based distributed observer, which has been proved to have the ability to approximate the control performance of the centralized control arbitrarily. The simulation results have verified the effectiveness of our theories.

\ifCLASSOPTIONcaptionsoff
  \newpage
\fi



%


\bibliographystyle{IEEEtran}
\bibliography{IEEEtran}

\end{document}